# Electromechanical response of saddle points in twisted hBN moiré superlattices

Stefano Chiodini[1]*, Giacomo Venturi[1,2], James Kerfoot[3], Jincan Zhang[3], Evgeny M. Alexeev[3], Takashi Taniguchi[4], Kenji Watanabe[5], Andrea C. Ferrari[3] and Antonio Ambrosio[1]*

[1]Center for Nano Science and Technology, Fondazione Istituto Italiano di Tecnologia, Via Rubattino 81, 20134, Milan, Italy

[2]Physics Department, Politecnico Milano, P.zza Leonardo Da Vinci 32, 20133 Milan, Italy

[3]Cambridge Graphene Centre, University of Cambridge, 9, JJ Thomson Avenue, Cambridge, CB3

[4]Center for Materials Nanoarchitectonics, National Institute for Materials Science, 1-1 Namiki, Tsukuba 305-0044, Japan

[5]Research Center for Functional Materials, National Institute for Materials Science, 1-1 Namiki, Tsukuba 305-0044, Japan

*corresponding authors

Email: stefano.chiodini@iit.it ; antonio.ambrosio@iit.it

**ABSTRACT:**

In twisted layered materials (t-LMs), an *inter*-layer rotation can break inversion symmetry and create an interfacial array of staggered out-of-plane polarization due to AB/BA stacking registries. This symmetry breaking can also trigger the formation of edge *in-plane* polarizations localized along the perimeter of AB/BA regions (*i.e.*, saddle point domains). However, a comprehensive experimental investigation of these features is still lacking. Here, we use piezo force microscopy to probe the electromechanical behavior of twisted hexagonal boron nitride (t-hBN). For a parallel stacking alignment of t-hBN, we reveal very narrow (width ~ 20 nm) saddle point polarizations, which we also measure in the anti-parallel configuration. These localized polarizations can still be found on a multiply-stacked t-hBN structure, determining the formation of a double moiré. We also visualize a t-hBN moiré superlattice in the topography maps with atomic force microscopy, related to the strain accumulated at the saddle point domains. Our findings imply that polarizations in t-hBN do not only point in the out-of-plane direction, but also show an in-plane component, giving rise to a much more complex 3D polarization field.

The detection and manipulation of electric,[1] magnetic[2] and valley polarizations[3] are key for device (*e.g.,* memories and logic circuits) performance optimizations.[4] As Moore's law approaches its physical limits,[4] the need for miniaturized nanoelectronics,[5] involving high-density integrated circuits and low power consumption[6] has triggered research into layered materials (LMs),[7,8] in order to reduce polarization domains from the 100 nm$^2$ scale down to the atomic scale.[5] Room temperature out-of-plane ferroelectricity is one of the main achievements of this approach, as it offers a wide range of technological applications, such as ultra-thin (few atomic sheets) non-volatile memories[9] and high-permittivity (compared to silicon dioxide) dielectrics.[9,10] However, only few suitable ferroelectric LMs have been identified so far, like $CuInP_2S_6$,[11] $In_2Se_3$,[12,13] $MoTe_2$[14] and $WTe_2$[15] in their 1T phase. In other widely studied LMs such as hexagonal boron nitride (hBN) and 2H-type transition metal dichalcogenides (TMDs), vertical polarizations cancel out,[16] due to the centrosymmetric lattice structure, which makes these crystals unpolarized. A possible way to engineer polarization in these LMs is to break the inversion symmetry by introducing a twist angle, $\theta_{TW}$, between top and bottom layers,[16,17,18] determining a periodic modulation of the interlayer atomic registry, *i.e.*, a moiré superlattice. In twisted hBN (t-hBN) structures the interfacial vertical alignment of the N and B atoms distorts the bonding $2p_z$ N electronic orbital,[17] locally creating an electric dipole moment that leads to a moiré superlattice characterized by adjacent domains with out-of-plane (OOP) polarizations pointing in opposite directions.[16,17,18,19] Refs. 18, 20, 21 predicted that in-plane (IP) polarizations can also appear at the domains' *edges* of t-hBN, resulting into 3D vectorial patterns with rich topological structures. Topology plays a key role in LMs, ranging from band theory to skyrmions in magnetic samples.[20] Topological domains in ferroelectrics[22,23,24] have received much attention owing to their novel functionalities, such as negative capacitance[25] and high-density (> 200 gigabits per square inch) information processing.[26] However, experimental proofs of the IP polarizations in t-LMs are limited to irregular t-hBN moiré patterns[27] or twisted double bilayer graphene samples.[28]

Here, we use piezo force microscopy (PFM) to reveal edge in-plane polarizations in t-hBN moiré superlattices for parallel and anti-parallel stacking. We find very sharp (width~20 nm) polarizations localized at the edges between different domains of the moiré pattern, called saddle points (SPs), not seen by other scanning probe microscopy (SPM) techniques, such as electrostatic force microscopy (EFM),[18, 29] amplitude-modulation kelvin probe force microscopy (AM-KPFM),[18] and tapping mode phase imaging.[30] We prove the universality of these SP features by systematically probing them for superlattices corresponding to different twist angles in the range 0.04° - 0.18°. We also explore samples consisting of three hBN stacks (*i.e.*, two twisted interfaces).[31] The superposition of SP polarizations arising at the two interfaces is still measurable by PFM, showing a double moiré. The possibility of interfacing multiple layer polarizations could pave the way for unconventional properties, such as modulations of moiré ferroelectric behaviors. We then extend the investigation about SP domains from their PFM electromechanical response to the inherent topographical modulation imparted to t-hBN at these locations. Using contact mode AFM, we reveal a t-hBN moiré superlattice in the topography map, related to the strain accumulated at the SP domains. This is the simplest way to visualize moiré patterns by SPM.

**Parallel stacking alignment in t-hBN**

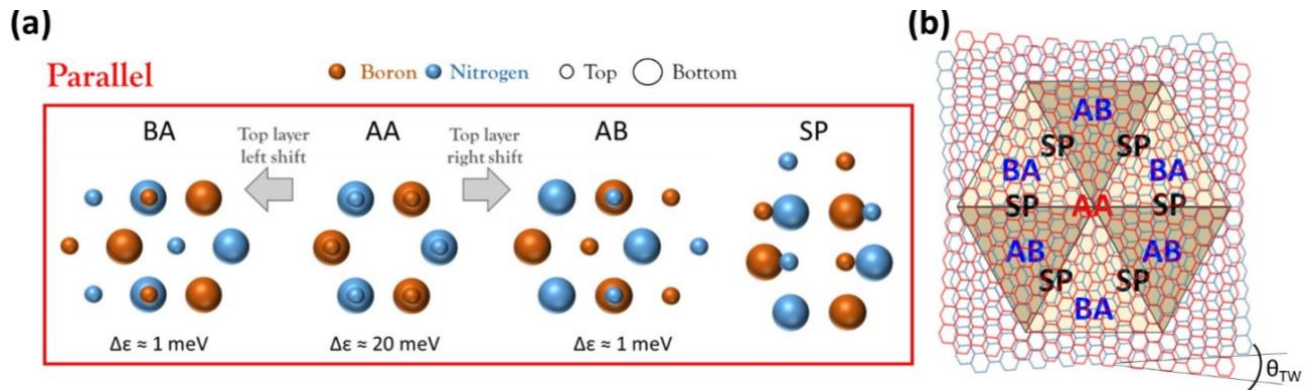

*Fig. 1.* *t-hBN parallel stacking configurations description. (a) Atomic registries corresponding to the 4 domains (AA, AB, BA, SP) typical of parallel stacking in a t-hBN interface (for the SP region we illustrate the average atomic registry). For AA, AB and BA configurations, the corresponding stacking energy per atom, Δε, relative to the naturally occurring AA' registry,[32] is reported below. B and N atoms of top (smaller circles) and bottom (larger circles) layers are sketched in maroon and*

*blue, respectively. (b) Representation of two hBN atomic layers, red and blue, (rigidly) stacked and twisted by a small angle $\theta_{TW} < 1°$. The superimposed internal drawing represents the typical 6 triangular shapes obtained after atomic relaxation, defining the moiré superlattice. The position of each of the 4 domains (AA, AB, BA, and SP) is shown.*

When two hBN layers are stacked together and twisted, the misalignment of the rotated atoms results in a periodic array of local stacking domains, *i.e.*, a moiré superlattice.[18] To rationalize the geometry of t-hBN stacking domains and their three-dimensional polarization network (IP and OOP), the hBN unit cell has to be considered. For t-hBN, two stacking alignments are possible, depending on the interfacial alignment of B and N atoms, *i.e.*, parallel and anti-parallel.[18, 21, 33] For parallel stacking, 4 different domains can be identified: AA, AB, BA, and SP. Their specific atomic registry is reported in Fig. 1a. The AA configuration is characterized by a full overlap between N (B) atoms of one layer and N (B) atoms of the twisted layer. In AB (BA) registry, the B (N) atoms in the top layer sit above the N (B) atoms in the bottom layer, while the N (B) atoms in the upper layer lay above the empty site at the center of the hexagonal cell of the lower layer. SP regions are between different domains, where the atomic registry changes from one domain to another.

The alternation of these 4 stacking regions forms the parallel moiré superlattice (where "parallel" refers to the stacking alignment) of t-hBN (Fig. 1b), according to a geometry which is set by a twist angle-dependent balance[16] between inter-layer interactions and intra-layer elasticity of the lattice, *i.e.*, the atomic relaxation. This is the driving force shaping the moiré domains geometry (triangular or hexagonal), mainly at $\theta_{TW} < 1°$, where atomic relaxation is more pronounced).[16, 17, 18, 34]

According to simulations,[32] AB and BA regions are energetically equivalent with a corresponding stacking energy (calculated with respect to the natural AA' stacking configuration)[32] $\Delta\varepsilon \sim 1$ meV (Fig. 1a) and, most importantly, energetically favorable with respect to the AA domain ($\Delta\varepsilon \sim 20$ meV), since the latter has pairs of N atoms atop of each other, resulting in an increased steric repulsion.[16] Hence, as shown in Fig. 1b, for parallel alignment at $\theta_{TW} < 1°$, AB/BA regions cover the majority of the moiré superlattice, with a triangular geometry,[18, 21, 33] while unfavorable AA domains are reduced to a smaller hexagonal coverage (Fig. 3f).[21, 27, 35, 36] Triangular AB/BA domains are characterized by oppositely oriented "ferroelectric-like" OOP polarizations,[18] whose properties were previously investigated *via* several SPM techniques such as EFM and KPFM.[16, 17, 18, 37] Only few experimental investigation have been carried out on IP polarizations arising from SP regions,[27, 28] with no extension to anti-parallel stacking, or any demonstration of their universal presence, independent from the twisting angle (at least inside the range investigated here).

In the supplementary information (SI), Section 1, we extend the description of the stacking domains to the t-hBN anti-parallel alignment.

**PFM of t-hBN parallel moiré superlattices**

t-hBN samples are prepared by exfoliating bulk hBN crystals, grown at high pressure and temperature in a barium boron nitride solvent,[38] onto Si+90nm SiO$_2$ by micromechanical cleavage (MC). In order to control $\theta_{TW}$, either large flakes (>50μm) selectively torn during transfer[39] or neighbouring hBN flakes cleaved from the same bulk crystal during MC[18] are identified by studying the orientation of their faceted edges using optical microscopy.[40] t-hBN samples with controlled *inter*-layer rotation are then fabricated using polycarbonate (PC) stamps.[41] First, a PC film on polydimethylsiloxane (PDMS) is brought into contact with the substrate with hBN flakes at 40 °C using a micromanipulator, so that the contact front between stamp and substrate covers part of one flake or one of two adjacent flakes exfoliated from the same flake on the tape. Stamps are then retracted, and the material in contact with the PC is picked up from the substrate. After picking up the first flake, a controlled $\theta_{TW}$ (± 0.01º), as determined by the resolution and wobble of the rotation stage, is applied by rotating the sample stage, before the flake on PC is aligned to the second one and brought into contact at 40 °C. The stamp is then retracted and the resulting t-hBN is picked up by PC. t-hBN is then transferred onto Si+285nm SiO$_2$ at 180°C, before the PC residue is removed by immersion in chloroform and then ethanol for 30 mins. While Si+90nm SiO$_2$ is used to facilitate the identification of hBN flakes,[42] Si+285nm SiO$_2$ is chosen for further characterization, such as gate dependent electrical measurements. Characterizations *via* Raman spectroscopy is discussed in Ref.30, as well as in Section 2 of the SI. We first consider a 2 nm-thick top hBN (~5 layers) on an 8 nm bottom hBN (>10 layers) on Si+285nm SiO$_2$ as a substrate. The two flakes are aligned at $\theta_{TW}$ ~ 0°. This sample is characterized by PFM, where a conductive tip is in contact with the surface, while an oscillating electrical bias is applied *via* the tip itself. Through PFM the electromechanical (EM) response can be measured.[43] We define the EM coupling as any effect that produces an electric field across the material in response to a surface or volume deformation and *vice versa* (*i.e.*, piezoelectric and inverse-piezoelectric effects,[43] respectively). Due to the inverse piezoelectric effect, an electromechanically active sample deforms under a bias, and this distortion couples with the cantilever motion, whose deflection is measured by the cantilever detection system (*i.e.*, the standard AFM optical lever system - such as for our microscope - or the more powerful interferometric displacement sensor).[43] More details in the SI, Section 3. The origin of this EM sample deformation can arise from two main effects, piezoelectricity (PZ) or flexoelectricity (FLX).[44] PZ allows conversion of mechanical *strain* into electric fields (and *vice versa*) and it arises only in non-centrosymmetric samples, *i.e.*, when a broken inversion symmetry is present.[45] FLX, instead, allows a material to polarize in response to a *strain gradient* (*i.e.*, mechanical bending), and, conversely, to bend in response to an electric field. Despite half a century of history, the latter has been less considered because of its expected weak strength at the macroscale.[44] However, at the nanoscale, FLX can compete with PZ, or be bigger.[44] FLX is a universal property of all materials, without any symmetry constraint.[46]

Fig. 2a, b show two representative PFM amplitude and phase images obtained on our t-hBN sample (topography reported in SI, Section 4). The moiré domains are characterized by narrow features at the edge of the triangular AB/BA regions (width ~ 20 nm, inset of Fig. 2b), which look the same in both trace and retrace maps (see Section 4, SI), ruling out any artifact. Fig. 2c is a zoom

of a (representative) triangular domain from the PFM amplitude map (Fig. 2a), where the 3 sides of the triangle (a, b, c) are highlighted.

Since we measure the EM response of the sample *via* vertical PFM, one could expect these features to emerge from OOP polarizations. However, IP polarizations detection through a vertical PFM technique is also possible, due to the buckling effect.[47, 48] This stems from cantilever buckling oscillations occurring when domains with IP polarization are aligned parallel to the long axis of the cantilever (see SI, Section 5). Based on this, we now prove the observed features to emerge from an IP contribution to the EM response. We expect these IP features to be measurable even in lateral PFM.[27]

If the buckling effect is relevant, we expect an angle-dependent PFM amplitude - *A* - signal,[27] *i.e.*, $A \approx P \cdot \sin\alpha = \vec{P_{i,y}}$, where $\vec{P}$ is the polarization vector, the label *i = a, b, c* refers to one of the sides of a triangle (defined as in Fig. 2c), and *α* is the angle between long cantilever y-axis and the triangle side under consideration. In Fig. 2e, the polarizations *P* are in red, while their y-component is in blue.

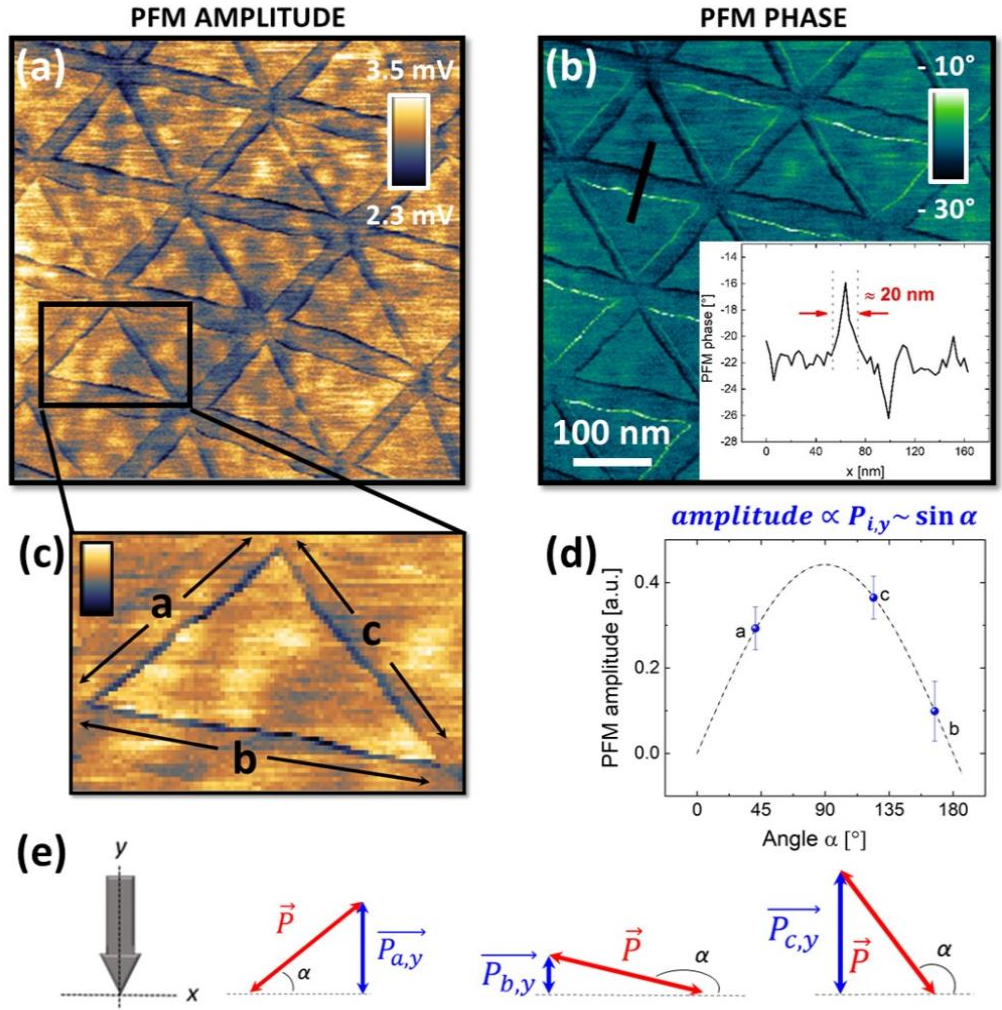

*Fig. 2.* IP polarizations measurement via vertical PFM in a t-hBN sample. (a, b) PFM amplitude and phase images of t-hBN. The inset of (b) is a PFM phase line profile highlighting a feature localized in a width~20 nm. (c) Zoom of a representative triangular domain in (a). The 3 triangle sides are labelled (a, b, c). (d) PFM amplitude as a function of the angle α between the triangle side (a, b, c) and cantilever main axis. The data are obtained averaging 7 triangular domains. The sinusoidal trend proves the PFM polarizations have an IP nature. Error bars correspond to standard deviations. (e) Vectorial decomposition of each polarization involved in the triangular shape in (c). Red: polarization vectors $\vec{P}$. Blue: $\vec{P_{i,y}}$ components along the y-axis (main cantilever axis) with i = a, b, c. α is measured with respect to the positive direction of the x-axis.
9

To corroborate this hypothesis, we report in Fig. 2d the measured average PFM amplitudes (along the triangular sides) as a function of $α$. The data fit a sinusoidal function, which can be considered the fingerprint of the IP nature of such polarizations localized along the triangular moiré domains SP.[27] Nevertheless, we cannot exclude the presence of edge OOP polarizations, as they could contribute to the PFM amplitude map with an α-independent offset (proportional to the effective piezoelectric coefficient, see SI, Eq. S2, S3) preserving the sinusoidal trend. We also measured a non-trivial EM signal in the PFM phase channel (Fig. 2b). Here, the adjacent sides of neighboring triangles offer staggered bright (~ -10°) or dark (~ -30°) phase values. Ref.20 suggested that IP polarizations localized at the SPs should curl around each triangle edge, with parallel polarizations along adjacent sides of neighboring triangles, which is not what we measure in Fig. 2b. Likely, our experimental result could emerge from an additional OOP polarization, still localized at the SPs. These OOP polarizations may emerge from saddle point PZ (and not FLX) as detailed in SI, Section 6.

Figs. 2a, b show that the internal area of the triangular domains does not offer any EM contrast between the adjacent triangular regions (AB/BA domains). Refs.17, 49 reported that nearby triangular AB/BA domains provide a PFM amplitude and phase contrast in the internal area. In our case, the different sample thicknesses could play a role. Indeed, here and in Ref.27 top and bottom flakes are not monolayers (1L) hBN. As a result of our larger flakes thicknesses, the vertical contrast between AB and BA polarizations could be hindered due to the intrinsic 3d nature of the atomic relaxation.[50]

We now extend the analysis of SP polarizations to moiré patterns arising from different twist angles. In order to consider them a general feature of such superlattices, they have to be present independently of $θ_{TW}$. The fact that $θ_{TW}$ may vary on a given sample is a consequence of the fabrication procedure, which does not allow for deterministic control, at the micron-scale, of $θ_{TW}$. Defects and fabrication residuals with unknown distribution over the sample areas can locally alter the twisted structure causing heterogeneous strain distributions and variations of $θ_{TW}$. Hence, when dealing with a t-LM at a specific $θ_{TW}$, we expect local deviations around the target value at the interface between top and bottom layers, which will also tune the moiré superlattice to a different periodicity ($Λ_m$, see

Fig. 3b). According to the theory of moiré superlattices, an inverse relation exists between $\Lambda_m$ and (the sine of) the twist angle, i.e., $\Lambda_m = (a/2)/\sin(\theta_{TW}/2)$,[28, 51] where in our case $a$ corresponds to the hBN lattice constant equal to 0.25 nm.[52] This formula is valid only under two assumptions: the layers are treated as rigid (i.e., atomic relaxation is neglected), and they are unaffected by strain. The latter constraint can be relaxed by considering the presence of strain as for Ref.51, as detailed in SI, Section 7. In our case, due to a negligible contribution emerging from strain (i.e., bubbles from fabrication or intrinsic strain of the individual flakes) applied to the top and bottom hBN flakes, the $\theta_{TW}$ variation with respect to an un-strained situation is calculated to be only ~5%. Regarding strain and $\theta_{TW}$ contributions coming from atomic relaxation, instead, the only way to account for these effects is to implement specific calculations within the density functional theory formalism as proposed in Refs53, 54. Nevertheless, such contribution is not expected to alter the periodicity of the moiré superlattice (as the AA domains' position is not affected by reconstruction)[55]. Therefore, our approach for $\theta_{TW}$ extraction remains valid.

Figs. 3a-e plot the PFM amplitude images obtained on different sample regions characterized by a decreasing (parallel) moiré pattern period. $\Lambda_m$ ranges from ~ 350 to ~ 80 nm, corresponding to an increasing estimated $\theta_{TW}$ from~0.04° to~0.18°. For each image, sharp features are present at the evolving SP regions, revealing the universality of this localized EM response of t-hBN.

Figs. 3a-e also allow us to evaluate the shape evolution of all atomic registries with $\theta_{TW}$. Their identification, from AB/BA domains (in blue) to AA (in red) and SP (in black) is presented in Fig. 3f. This image proves AA domains to have a hexagonal shape, as theoretically expected[21, 36], but, thus far, not observed experimentally, to the best of our knowledge. Going from Fig. 3a to e, there is a decreasing coverage of triangular AB/BA regions, in favour of AA hexagonal domains, progressively growing in size. To quantify this evolution, we first need to define a super-hexagon for each PFM image of Figs. 3a-e. This acts as an effective "unit-cell" for the moiré superlattice and encloses three AB and three BA triangular domains. E.g., the super-hexagon corresponding to Fig. 3d is highlighted by the white dashed line in Fig. 3f. Second, for each PFM amplitude image, we can obtain the areal

coverage[56] for AA, AB/BA, SP regions confined inside the corresponding super-hexagon. These areas can be normalized dividing by the total coverage of the super-hexagon itself, defining what we have called in Fig. 3g "relative stacking area". The $\theta_{TW}$-dependent evolution of the relative stacking area for AA, AB/BA, SP regions is illustrated in Fig. 3g. Different trends are observed: while the AA relative coverage increases with $\theta_{TW}$, (red data), the AB/BA behavior (blue data) decreases. This is consistent with a $\theta_{TW}$-dependent atomic relaxation, progressively decreasing the relative area covered by AB/BA triangular domains at larger angles, favoring AA regions instead.[36]

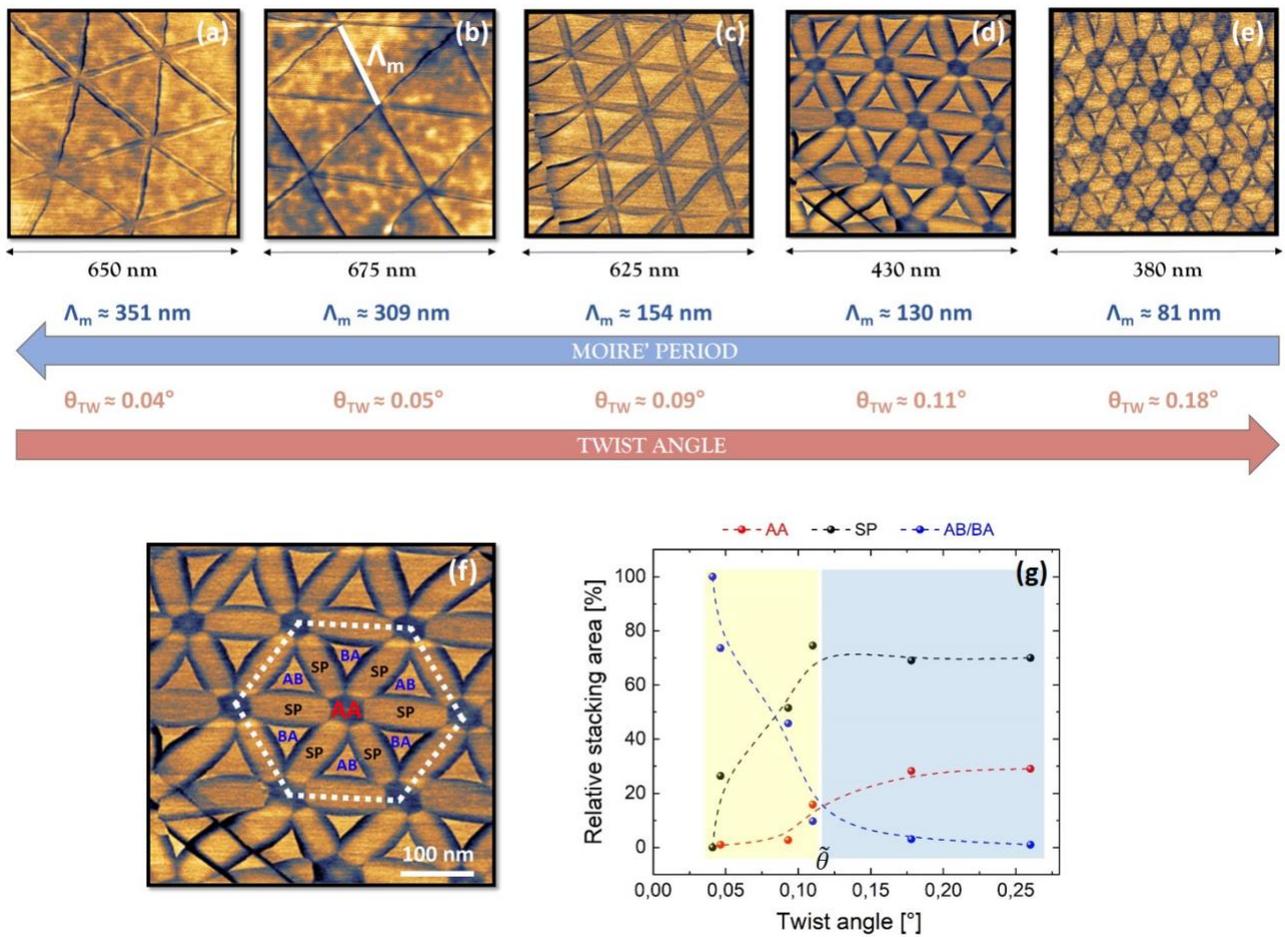

*Fig. 3. Parallel stacking domains evolution with increasing $\theta_{TW}$. (a-e) 5 moiré patterns characterized by a different moiré period ($\Lambda_m$) corresponding to an increasing $\theta_{TW}$ between top and bottom hBN. At the bottom of each image, the scan size is reported. The big blue and red arrows show the evolution of $\Lambda_m$ and $\theta_{TW}$, respectively. For the determination of $\theta_{TW}$, see the SI, Section 7, while $\Lambda_m$ is experimentally determined as the average distance between AA domains.[52] Amplitude scale bar: (a)*

*2.7- 6.6 mV, (b) 1.3- 2.3 mV, (c) 6.3 - 9.9 mV, (d) 8.5- 14.5 mV, (e) 19 - 24 mV. (a, b) are for a t-hBN with top flake thickness~2 nm, bottom flake thickness~8 nm, $\theta_{TW}$ ~ 0°. (c, d, e) are for a different t-hBN with top layer thickness~4.5nm, bottom flake thickness~40 nm, $\theta_{TW}$ ~0.2°. (f) PFM amplitude map (equivalent to Fig. 3d), with all stacking domains identified on the surface (AB/BA, AA, SP). The white dashed line represents the moiré super-hexagonal shape used in the relative stacking area quantification of (g). (g) Relative (%) stacking area evolution with increasing $\theta_{TW}$ for AA (red), SP (black) and AB/BA (blue) regions. Following the SP trend, 2 regions can be highlighted: a first (yellow), for $\theta_{TW} < \tilde{\theta}$ ~ 0.1°, where the SP relative area is increasing with $\theta_{TW}$, and a second (blue) where this trend saturates reaching a plateau for $\theta_{TW} > \tilde{\theta}$ ~0.1°. The last point on the right of the plot ($\theta_{TW}$ ~ 0.25°) is obtained on the additional PFM image in Fig. 5b, corresponding to $\Lambda_m$~55 nm.*

There is a point where the relative coverages of AB/BA and AA domains balance, marking the boundary between two reconstruction regimes where energetically unfavorable domains take over. This happens at $\theta_{TW} = \tilde{\theta}$ ~ 0.10°, further confirmed by following the SP relative area evolution (black data). For θ < $\tilde{\theta}$ we observe a linear SP trend, which then reaches a plateau for θ >$\tilde{\theta}$. A similar trend was reported for t-BLG,[35] where $\tilde{\theta}$ assumes a higher relevance as it marks the appearance of flat bands, *via* 4D-scanning transmission electron microscopy (4D-STEM),[35] a technique which shares with PFM the ability to measure lattice strain. However, PFM does not require any dedicated vacuum environment, therefore simplifying the strain mapping. Hence, we believe PFM could be employed for the identification of correlated electronic states in t-LMs.

The analogous evolution of the EM response of moiré superlattices for the anti-parallel alignment probed at two different $\theta_{TW}$ is presented in Section 8, SI.

**PFM of t-hBN anti-parallel moiré superlattices**

We further generalize the relevance of SP polarizations by experimentally revealing their presence also for t-hBN anti-parallel stacking alignments. To access this additional interfacial alignment, we exploit the topography of a t-hBN sample with top flake thickness~4.5 nm, bottom flake thickness ~ 40 nm, $\theta_{TW}$ ~ 0.2°, offering a 1L step underneath the top flake, Fig. 4a (see Fig. 4e for a sketch of the sample structure). Since multilayer hBN (ML-hBN) such as the bottom flake, has a natural AA' stacking,[18] the addition of a 1L step, would produce a rotation of 180° with respect to the underlying structure, determining a parallel to anti-parallel stacking transition (Fig. 4e).[18] Fig. 4a confirms the step to correspond to 1L of hBN,~0.3 nm.[18] Fig. 4b plots the related PFM phase (see Section 9, SI, for the corresponding PFM amplitude map). While triangular AB/BA domains are visible on the top-right part of these three images (parallel interfacial stacking), the bottom-left region, corresponding to the 1L-hBN addition, shows hexagonal structures typical of *anti-parallel* stacking, with features localized at the SP domains (see also Fig. S8d). Figs.4c, d show the corresponding AM-KPFM and phase-imaging maps of the same region. While all three SPM techniques allow the visualization of AB/BA triangular domains, PFM is the only approach capable of visualizing anti-parallel stacking.

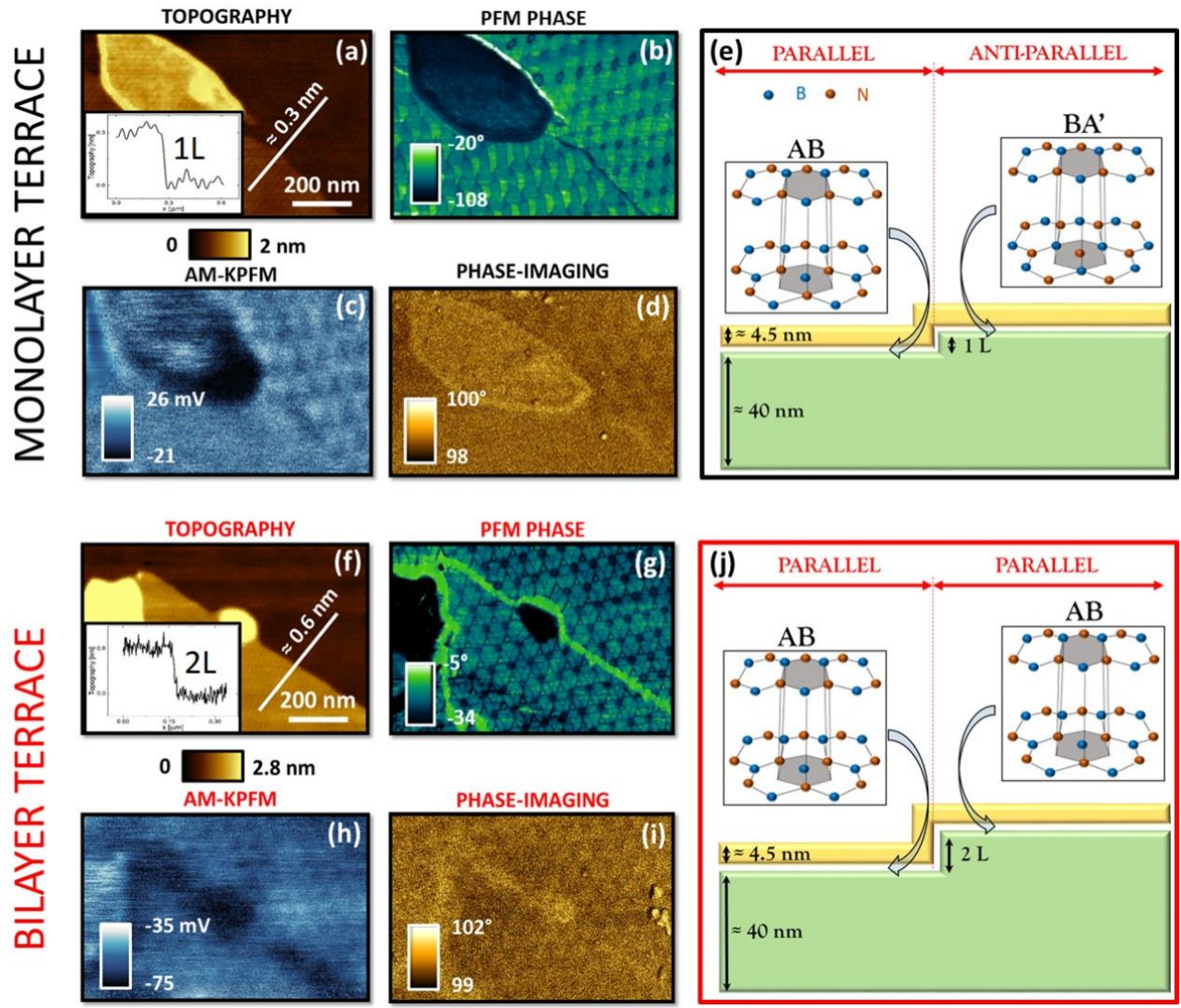

***Fig. 4.*** *t-hBN parallel and anti-parallel stacking domains measured with different SPM techniques. (a-e): Parallel to anti-parallel alignment transition induced by a 1L topographical step~0.3 nm on a t-hBN sample with top flake thickness~4.5nm, bottom flake thickness~40 nm, $\theta_{TW}$~0.2°. (a) AFM topography of 1L step. (b-d) PFM phase, AM-KPFM, and phase-imaging maps in the same region of (a). (e) Schematic sample structure corresponding to Figs.4a-d. On the top part a possible stacking transition is shown from parallel AB to anti-parallel BA' lattice registry. (f-j): Parallel to parallel alignment transition due to a 2L-hBN topographical step~0.6 nm. (f) AFM topography of 2L-hBN step. (g-i) corresponding PFM phase, AM-KPFM, and phase-imaging channels. (j) Schematic sample structure corresponding to Figs.4f-i. The top part of panel (j) sketches a parallel stacking domain (AB) on both sides of the 2L-hBN step. The thicknesses of the flakes and steps are not to scale.*

To validate this further, we focus on a different region of the same sample offering a topographical 2L-hBN step ~ 0.6nm (see Fig. 4f for the topography and Fig. 4j for the sample structure). If a 1L-hBN step is responsible for a 180° rotation, it follows that a 2L-hBN step does not induce any parallel to anti-parallel stacking transition. Figs. 4g-i show the PFM phase (PFM amplitude image shown in Section 9, SI), AM-KPFM, and phase-imaging maps of the same region, addressing a parallel stacking on both sides of the 2L-hBN step.

These SP polarizations can be identified only by probing the t-hBN EM properties by means of PFM, suggesting a different imaging mechanism for the three SPM techniques used in Fig. 4. AM-KPFM is non-contact[18] and measures local tip-sample electrostatic interactions, allowing to directly access the contact potential difference through a double-pass technique[18] (increasing the scanning time with respect to the single-pass PFM). According to Fig. 4c, this approach yields a moiré pattern only for the parallel stacking. We attribute this observation to the dependence of the contact potential difference on the local dipoles. There are much more pronounced in the triangular parallel moiré superlattice.[18] Anti-parallel domains offer to the surface an (OOP) polarization ~3 orders of magnitudes smaller,[18] typically below the sensitivity of AM-KPFM (similar considerations can be applied to EFM).[18] Regarding phase-imaging, when performed in the attractive regime as in Fig. 4d,[30] the probing mechanism relies on the van der Waals (Debye-like, *i.e.*, dipole-dipole) tip-sample interaction which, once again, is stronger for a parallel alignment.[18]

Hence, PFM is a convenient SPM technique for visualizing *both* parallel and anti-parallel stacking (Fig. 4b), since it relies on the EM, rather than electrostatic interaction between tip and sample.

**Double-moiré and moiré-modulated topography**

There is an increasing interest in graphene t-2L-LMs and t-ML-LMs, due to their emerging superconducting[57, 58, 59] and correlation insulating behaviors[60, 61, 62] and in t-ML-TMD where cumulative polarizations have been recently measured by KPFM.[31] We then consider a specific region of our t-hBN sample where the fabrication process introduced an additional layer, with an uncontrolled orientation relative to the underneath hBN. Effectively, this turns the area into a t-ML-hBN sample. Fig. 5a is a PFM amplitude map obtained in this zone which is related to a topographical step (~ 4 nm, see Fig. 5c), separating 2 different regions. The top-left part of the image involves a moiré superlattice made of big ($\Lambda_m$ ~ 300 nm) triangular AB/BA domains. The bottom-right part of the image has two overlapped textures: a first superlattice which follows the previously discussed pattern (made of big triangular domains), plus a second finer superlattice whose tiny details can be visualized through a high-resolution PFM amplitude map, see Fig. 5b. Fig. 5d is an AM-KPFM map of the same region of Fig. 5a. Only the first superlattice can be distinguished, probably due to a weaker IP signal and/or a limited spatial resolution of AM-KPFM.[63]

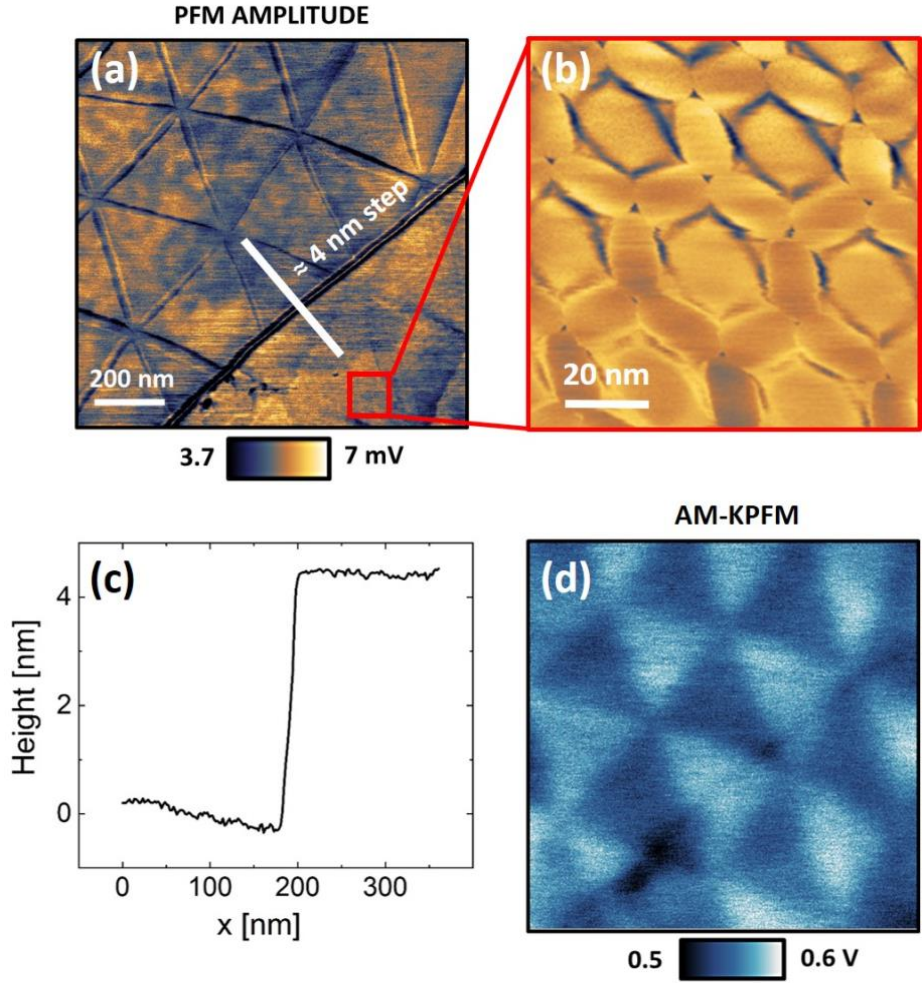

*Fig. 5. Double-moiré in multiply-stacked t-hBN structure measured via PFM. (a) PFM amplitude map showing a double-moiré in the bottom-right corner. (b) Zoom of red square in (a). (c) Height profile across the 4nm step highlighted in panel (a). (d) AM-KPFM map of the same area of Fig. 5a.*

From Fig. 5b, we derive $\Lambda_m \sim 50$ nm, smaller than the typical dimension of the first superlattice ($\Lambda_m \sim 300$ nm). There is a completely different geometry of the fine pattern, mainly characterized by hexagonal structures, corresponding to central AA stacking domains. The rounded areas surrounding AA regions may be SP domains, with AB/BA regions limited to very small (but still visible) triangular domains. Considering the ML-hBN structure in this region (schematic in Section 10, SI), we ascribe this PFM experimental observation to the presence of a double-moiré (in the bottom-right part of Fig. 5a), emerging from three t-hBN stacks. In this region, despite a ~ 4 nm ML-hBN topographical step, the SP features of the larger pattern are still measurable by PFM, providing evidence for a robust

PFM signal arising from SP polarizations. This observation is also of practical importance, since it simplifies access to double moiré superlattices without the need of using 1L- or few layers-crystals, which can be experimentally challenging.

This finding could trigger interesting theoretical and experimental investigations based on moiré-moiré interactions. E.G. the OOP polarizations in AB/BA domains localized on both superlattices could couple, providing a modulation of the ferroelectric behavior of t-hBN.

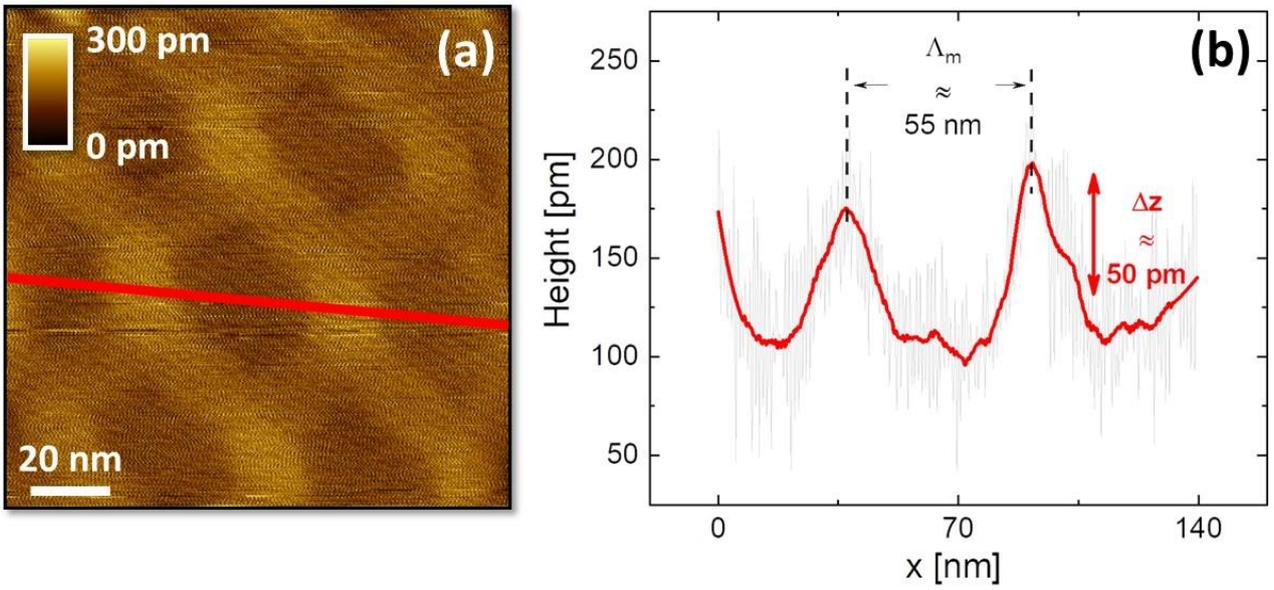

*Fig. 6. Topographic moiré measured via PFM. (a) PFM topography (with applied bias, $V_{dc} = 0V$, $V_{ac} = 2V$) in the same region of Fig. 5b. (b) Topographical line profile related to the red line of (a), showing $\Lambda_m \sim 55$ nm and $\Delta z \sim 50$ pm.*

We now consider a route to visualize moiré superlattices already in the topography of a t-LM. At small $\theta_{TW}$ twist angles, when atomic relaxation takes place, the resulting strain in moiré patterns is expected to evolve from zero to higher values as long as we increase $\theta_{TW}$.[28] Since, locally, the strain accumulates mostly at the SPs,[35] this could trigger a measurable (above the image noise level) SP topographical moiré corrugation, depending on $\theta_{TW}$.

Fig. 6a visualizes the contact mode topography associated with the PFM image of Fig. 5b. The surface is not flat, and a moiré superlattice is visible following the features of Fig. 5b. To rule out any PFM voltage-related artifacts, we re-scan the area setting the PFM bias (DC and AC) to zero (see Section 11 of the SI). This yields the same topography image, confirming the intrinsic morphological nature of the contrast. In Fig. 6b, a topographical profile is visualized (following the red line of Fig. 6a), showing $\Lambda_m$ ~ 50 nm and $\Delta z$ ~ 50 pm, localized along the SP domains, *i.e.,* the regions where strain is mostly accumulated.[35] No analogous features are present in the PFM topography of another t-hBN region characterized by a bigger triangular moiré pattern ($\Lambda_m$ ~ *250 nm*, Fig. S10c, d). This is in agreement with our interpretation: at higher $\Lambda_m$, *i.e.*, smaller $\theta_{TW}$, we expect a less pronounced accumulation of strain at the SPs, determining a smaller modulation of the hBN top surface, eventually hindered by the image noise level.

The possibility to visualize a moiré superlattice in the topography of a t-LM was reported, e.g., by scanning tunneling microscopy (STM) on $1L-MoS_2/1L-WSe_2$.[64] However, PFM does not need any dedicated vacuum environment, simplifying the measurement and reducing the experimental time.

**Conclusions**

We used PFM to probe the local electromechanical properties of t-hBN, showing the formation of in-plane polarizations at the edges of the stacking domains (saddle points) of both parallel and anti-parallel moiré superlattices. We explained the in-plane and out-of-plane origin of these saddle point polarizations, proving their universality by evaluating moiré superlattices for a range of twist angles. The relevance of these saddle point polarizations was extended by measuring them also in a double-moiré emerging from the relative twisting of three hBN stacks involving two interfaces. The strain typically localized at the saddle point regions allowed us to measure a moiré superlattice in the topography, *via* standard contact mode atomic force microscopy.

Our work unveils a richer polarization (in- and out-of-plane) network in t-hBN, whose spatial distribution can be tuned by the twist angle, as opposed to conventional ferroelectric materials,[17] where the polarization domains are determined by the crystal structure. This complex 3D vectorial polarization pattern could trigger interesting topological investigations,[20] related to negative capacitance[25] or high-density information processing[26], but also provide new insights for exploring unconventional behaviours in t-LMs. In this context, the experimental observation of a double-moiré is important, due to the properties observed in t-ML-graphene, where superconducting[57, 58, 59] and correlation insulating properties[61, 62] have been found, and in t-ML-TMD, where cumulative polarizations were measured.[31] Similarly, the emergence of a double-moiré in t-hBN involving both IP and OOP polarizations, could pave the way for moiré ferroelectricity modulations *via* multi-stacking.[37] 37

The observation of a topographical moiré superlattice by contact mode AFM, instead, provides a tool for moiré superlattices visualizations free from complex experimental set up and/or sample preparations. This could trigger numerical investigations aiming for a full understanding of the twist angle dependent strain distribution along moiré pattern saddle points.

The ability of PFM to image *both* parallel and anti-parallel t-hBN alignments, with high spatial resolution (about 10 nm), not possible with other SPM techniques, confirms it as a very powerful technique to study moiré superlattices in t-LMs.

## METHODS

*Atomic Force Microscopy*

AFM measurements are performed at ~25 °C (RH~40%), in air, using a Multimode 8 (Bruker) AFM microscope. To avoid damaging the tip, the cantilevers sensitivity calibration procedure is performed at the end of the experiments. The deflection sensitivity is obtained by performing 10 force-distance curves on mica (without changing the laser spot position onto the cantilever) and calculating the average inverse slope in the contact region. An average value of 68 nm·V$^{-1}$ is found. SCANASYST FLUID cantilevers (Bruker, $k$~0.7 N·m$^{-1}$, $f$~150 kHz) are used for phase-imaging, while ASYELEC.01-R2 cantilevers (Asylum Research, $k$~2.8 N·m$^{-1}$, $f$~75 kHz) for all the PFM and AM-KPFM images. The phase-imaging typical parameters are free amplitude $A_0$~8 nm, set-point~7 nm. The attractive phase values in this work are reported following the Asylum Research convention[21]. For PFM, we use a set-point~5 nm, with typical contact resonance $f_{CR} \approx 330$ kHz, and an AC sample bias amplitude $V_{ac}$=2 V ($V_{dc} = 0$, referring to Eq.S1), with grounded tip. In AM-KPFM, the images are acquired with $A_0$~20 nm, set-point~5 nm, lift height~2 nm, lift driving voltage~2 V, sample grounded. All AFM images are obtained at a typical scan rate of 0.8 Hz and analyzed in Gwyddion.[56] AM-KPFM maps are flattened together with a second order polynomial correction to enhance the moiré contrast between AB and BA triangular domains.

# AUTHOR INFORMATION

## Corresponding Authors

**Stefano Chiodini** - Center for Nano Science and Technology, Fondazione Istituto Italiano di Tecnologia, Via Rubattino 81, 20134, Milan, Italy;

Email: Stefano.chiodini@iit.it

**Antonio Ambrosio** - Center for Nano Science and Technology, Fondazione Istituto Italiano di Tecnologia, Via Rubattino 81, 20134, Milan, Italy;

Email: antonio.ambrosio@iit.it

## Author Contributions

S.C. developed the investigation approach and performed and analyzed the AFM measurements. G.V. performed the angle extraction from the PFM data. J.Z., E.M.A., and A.C.F. prepared and characterized all the samples. T.T. and K.W. provided the bulk crystals. S.C., G.V., A.C.F., and A.A. wrote the paper with contributions from the other authors. All the authors discussed the experimental data and the paper content. A.C.F. and A.A. coordinated the research activities.

## ACKNOWLEDGEMENTS

We acknowledge funding from ERC grants "METAmorphoses", grant agreement no. 817794, Fondazione Cariplo, grant n° 2019-3923, EU Graphene Flagship, ERC grants Hetero2D, GIPT, EU grants Graph-X,

CHARM, EPSRC grants EP/K01711X/1, EP/K017144/1, EP/N010345/1, EP/L016087/1, EP/V000055/1, EP/X015742/1.

**Competing financial interests**

Authors declare to have no competing financial interests.

*Supplementary Information*

# Electromechanical response of saddle points in twisted hBN moiré superlattices

Stefano Chiodini[1], Giacomo Venturi[1, 2], James Kerfoot[3], Jincan Zhang[3], Evgeny M. Alexeev[3], Takashi Taniguchi[4], Kenji Watanabe[5], Andrea C. Ferrari[3] and Antonio Ambrosio[1]

[1]Center for Nano Science and Technology, Fondazione Istituto Italiano di Tecnologia, Via Rubattino 81, 20134, Milan, Italy

[2]Physics Department, Politecnico Milano, P.zza Leonardo Da Vinci 32, 20133 Milan, Italy

[3]Cambridge Graphene Centre, University of Cambridge, 9, JJ Thomson Avenue, Cambridge, CB3

[4]Center for Materials Nanoarchitectonics, National Institute for Materials Science, 1-1 Namiki, Tsukuba 305-0044, Japan

[5]Research Center for Functional Materials, National Institute for Materials Science, 1-1 Namiki, Tsukuba 305-0044, Japan

**Section 1: Anti-parallel stacking alignments in twisted hexagonal boron nitride (t-hBN)**

In the anti-parallel stacking alignment AA', AB' and BA' lattice registries can form (Fig. S1) together with saddle points regions (Fig. 1a). AA' domains are characterized by the lowest stacking energy,[1] $\Delta\varepsilon$, with AB' and BA' domains described by higher (and different) energy values.[1] Hence, for the anti-parallel stacking, atomic relaxation favors[2, 3, 4] AA' regions, with AB' and BA' domains having a minor coverage along the moiré superlattice. AA' stacking has a hexagonal shape,[2, 3] while AB'/BA' domains are more triangular and develop an out-of-plane (OOP) charge density which is three orders of magnitudes weaker than in parallel stacking (AB/BA domains) [2, 3].

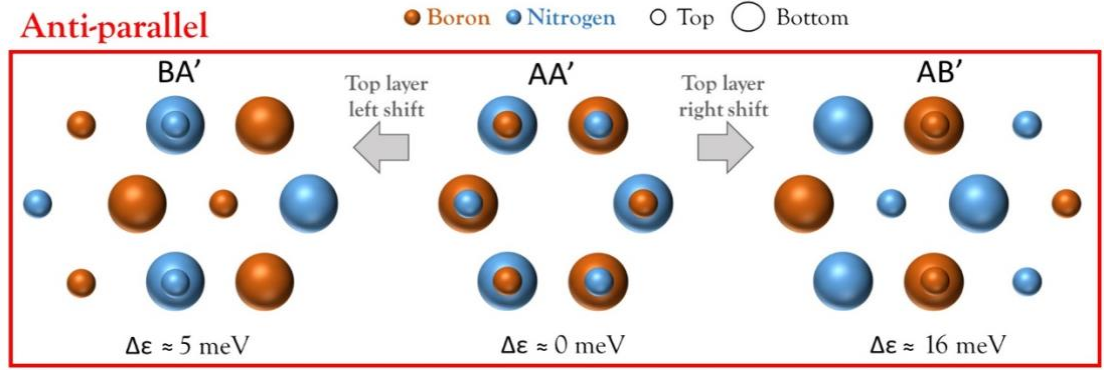

*Fig. S1.* *Stacking domains for anti-parallel alignment in t-hBN. Starting from the natural configuration of hBN, AA' (centre of the figure), two other stacking can exist with different Δε with respect to AA' domains: BA' (left) or AB' (right). B and N atoms of top (smaller circles) and bottom (larger circles) layers are sketched in maroon and blue, respectively.*

**Section 2: Raman characterisation of the t-hBN sample**

To monitor the quality and track any induced strain and disorder within the hBN heterostructures throughout fabrication, Raman spectroscopy is performed using a Horiba LabRAM Evolution at 514 nm, with an 1800 l/mm grating and volume Bragg filters with a ~5 cm$^{-1}$ cut-off frequency and a 100x objective (NA: 0.9). Errors associated with the peak position and full width at half maximum (FWHM) are systematically explored by taking into account fitting error, spectrometer registry and statistical and transient variations[13]. In Ref. 5 we characterised the Raman spectra of the t-hBN sample referred to Fig. 2, Fig. 3a, b, 5, 6. Raman spectroscopy is used to confirm the thickness of top and bottom flakes *via* the position of the shear mode (C). Pos(C) can be used to determine the layer number (N), for N > 2, [14,15,16] with N = 5 extracted for the top hBN flake and N>10 for the bottom flake. The flake thicknesses are ~2 and ~8 nm by atomic force microscopy (AFM), in agreement with Raman spectroscopy.

Raman spectroscopy is also performed on hBN on native Si + SiO$_2$, as for Figs.3c-e, 4. The thickness of the bottom flake is ~43.7nm, while the top flake varies from 1L-hBN to up to 6.5 nm, as confirmed by AFM. Fig. S2a plots the Raman spectra of one of the bottom hBN flake. Owing to the differing intensity of reflections from the native Si versus Si+90nm SiO$_2$ of Ref. 5, we observe a significant increase in the ultra-low frequency background, attributed to reflected light from the 514 nm laser. The background is subtracted by fitting the spectra to an exponential function $A_1 e^{\frac{x-x_0}{c_1}} + A_2 e^{\frac{x-x_0}{c_2}}$, as shown by a red line in Fig.S2a, to give the spectra in grey in Fig. S2a. Figs. S2b, c plot the Raman spectra of the bottom hBN flake, t-hBN and the starting bulk hBN (B-hBN). For the bottom 43.7 nm hBN, t-hBN on Si and the starting B-hBN: Pos(C) = 52.4 ± 0.14 cm$^{-1}$, with FWHM(C) = 1 cm$^{-1}$ ± 0.2 cm$^{-1}$. Pos(C) can be used to determine N, for N > 2, [14,15,16]

$$\text{Pos(C)} = \frac{1}{\sqrt{2}\pi c}\sqrt{\frac{\alpha_\perp}{\mu}}\sqrt{1+\cos\left(\frac{\pi}{N}\right)} \tag{S6}$$

With c the speed of light in cm s$^{-1}$, μ = 6.9 × 10$^{-27}$ kg Å$^{-2}$ the mass of one layer per unit area and α$_\perp$ the interlayer coupling. [13, 14, 15] From Eq. (S6), we estimate N > 10 for the bottom hBN and t-hBN. Fig. S2c gives Pos($E_{2g}$) = 1365.2 ± 0.2 cm$^{-1}$ with FWHM($E_{2g}$) = 7.4 ± 0.2 cm$^{-1}$ for 43.7 nm, t-hBN, and B-hBN, which implies that the strain across the bottom flake and the t-hBN is <0.007 % based on the $E_{2g}$ shift rate[17]. Notably, this strain evaluation should not be considered perfectly accurate due to a diffraction limited Raman spot which necessarily takes into account many t-hBN moiré domains.

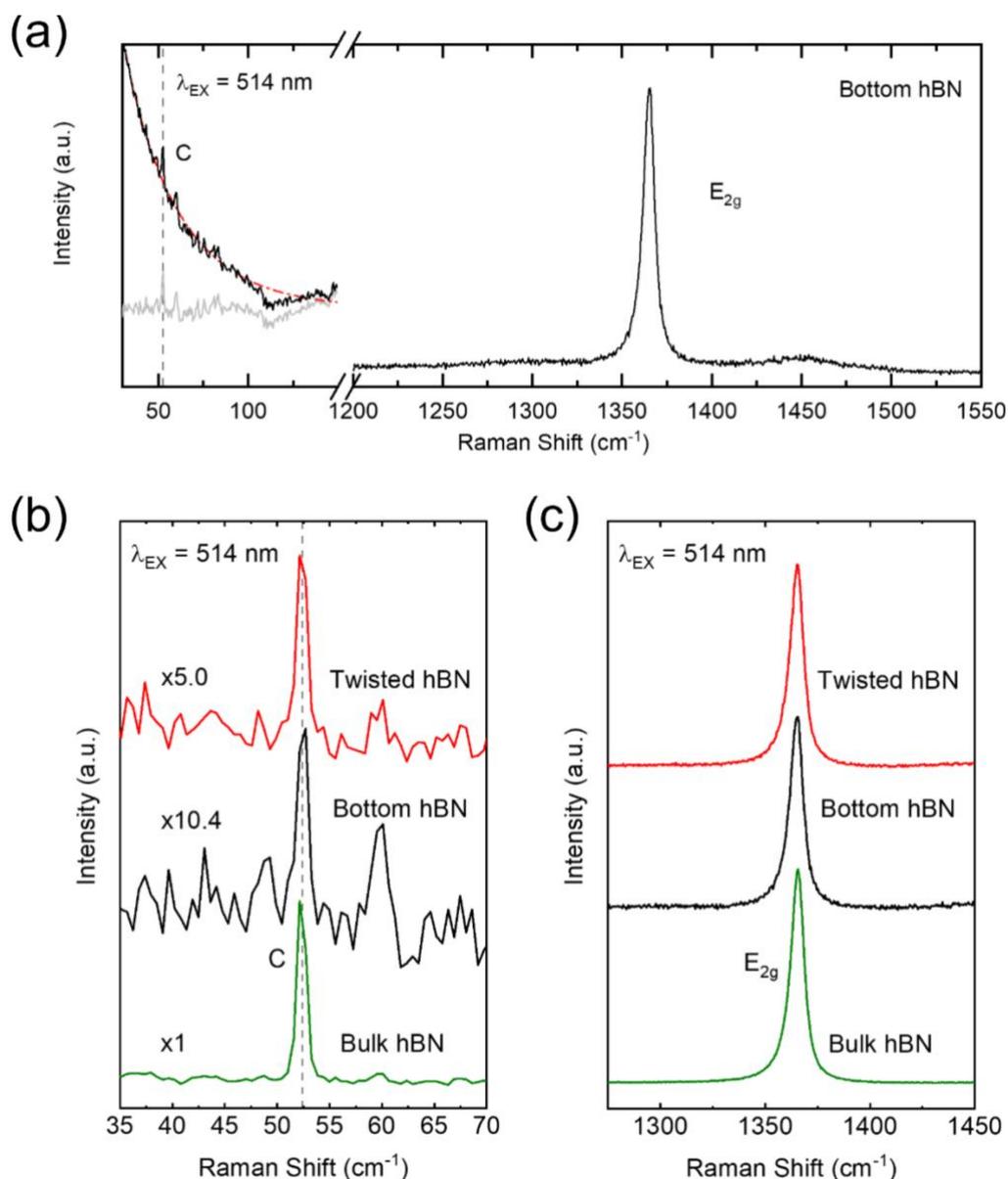

***Figure S2.*** *(a) Raman spectra of bottom 43.7 nm hBN on native Si is shown in black alongside the fit to the ultra-low frequency background, red. (b) Corrected low frequency spectrum. Low and (c) high-frequency Raman spectra of t-hBN (red), 43.7nm hBN (black) on Si and B-hBN (green). The spectra in (b, c) are normalised relative to I(C) and I($E_{2g}$).*

**Section 3: Resonance-enhanced vertical piezo force microscopy (PFM)**

Through standard PFM, or the more advanced dual AC resonance tracking[5] and band excitation[6] techniques, a plethora of different samples have been studied, from ferroelectric materials[7] to bio-samples,[8] perovskites,[9,10] and layered material (LM) moiré superlattices from graphene,[11,12] to transition metal dichalcogenides (TMDs)[13] and hBN.[12]

We visualize the electromechanical (EM) response of t-hBN samples by applying resonance-enhanced vertical PFM to circumvent the issue of materials with a weak (smaller than the cantilever thermal noise) EM response. This technique relies on electrically driving (see below) the cantilever at the contact resonance frequency,[14] in order to boost the signal-to-noise-ratio through resonance amplification. It provides better results for flat surfaces (such as those of LMs) where, due to a small roughness (or root-mean-square (RMS), see Fig. S3a) the contact resonance frequency can be considered nearly constant.

When the sample is piezoelectric (PZ), we expect a non-zero PZ tensor relating the applied potential to the sample deformation (referred to as the "inverse piezo effect")[9], which results in an induced polarization. The most relevant component of this tensor for typical PFM studies is the effective piezoelectric coefficient $d_{33,\text{eff}}$, considered as an effective parameter since different EM contributions can determine its magnitude (not only PZ, but also other terms such as electrostatics), together with the cantilever dynamics.[15] In PFM measurements, the voltage applied to the tip (or sample) is:[16]

$$V_{tip} = V_{dc} + V_{ac} \sin(2\pi f t) \tag{S1}$$

causing a sample deformation, traced by the cantilever periodic motion in time ($t$):[16]

$$z = z_{dc} + A(f, V_{dc}, V_{ac}) \sin(2\pi f t + \varphi), \tag{S2}$$

where $A$ is the PFM amplitude and $\varphi$ the phase, measured with demodulating lock-in. In resonance-enhanced PFM, $f$ is equal to the contact resonance frequency ($f_{CR}$).[14] Thus, Eq. (S2) becomes:

$$z = d_{33,eff} V_{dc} + d_{33,eff} V_{ac} Q \sin(2\pi f_{CR} t + \varphi) \tag{S3}$$

where $Q$ is the quality factor of the amplitude resonance curve at $f_{CR}$.[14]

When the sample is EM active, the PFM amplitude contains information on the magnitude of the effective PZ coefficient,[9] while the PFM phase allows the determination of the polarization direction.[9]

$\varphi = 0°$ means that the polarization is parallel to the applied electric field, while for $\varphi = 180°$ the local sample polarization and the electric field are anti-parallel.

**Section 4: Fig. 2 full set of data (topography, trace and retrace of PFM amplitude and phase)**

Fig. S3a provides the topography corresponding to Fig. 2. The *z*-scale bar and the RMS confirm the flatness of the surface. Figs. S3b-e compare the PFM amplitude and phase channels acquired during a full pass of the AFM tip along each line of the image (trace in panels (b, c) and retrace in panels (d, e)). The amplitude and phase signals at the borders of the triangle domains are not an artifact, as the images do overlap.

Despite the quality of Figs. S3b-e, there are two unexpected observations regarding the amplitude and phase values. First, there is a phase contrast<180°, unlike what is expected for opposite

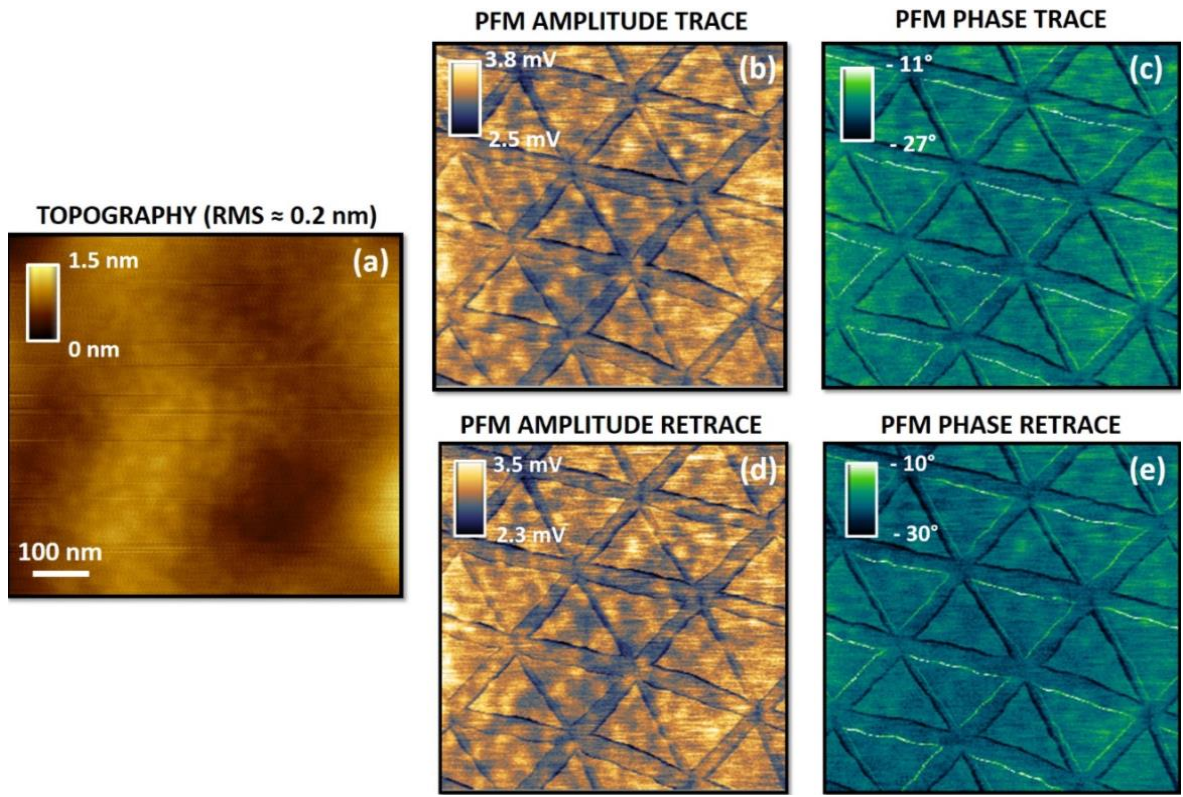

***Fig. S3.*** *(a) Topography corresponding to PFM amplitude and phase channels of Fig.2. (b, c) Trace PFM amplitude and phase channels. (d, e) Retrace PFM amplitude and phase maps.*

in-plane (IP) polarization directions. Second, the amplitude values do not always have a zero-minimum.[11, 15, 17]

A similar behavior was reported in Refs. 13, 19. This inconsistency may rise from a constant background imposed on top of the *real* signal[11]. In the most general case, this background emerges from the interplay between several phenomena, *e.g.* piezoelectricity, electrostatics, electrochemistry, and Joule heating, which cannot be controlled.

Regarding electrostatic contributions coming from the body of the cantilever, the electrostatic blind spot (ESBS) technique[15] provides a convenient way to reduce it. However, this approach is effective only for off-resonance PFM. Due to a weak EM response of the t-hBN sample, our measurements are performed on-resonance. Nonetheless, this electrostatic contribution is not dependent on the relative orientation between cantilever main axis and side of any triangular moiré domain (see discussion for Fig. 2 of the main text). Therefore, if it provides an additional PFM signal, it is expected to correspond to an offset, common to each moiré domain.

**Section 5: Buckling effect**

The buckling effect[18, 19] stems from the cantilever buckling oscillations that take place when domains with IP polarization are aligned parallel to the long axis of the cantilever itself.Fig. S4 shows the three possible orientations between the cantilever long axis and the sample polarizations. Fig. S4a reports the most standard case where OOP polarization regions are probed. Accordingly, the cantilever will bend vertically (the tip is displaced up and down, following the sample) providing a vertical signal on the photodiode. If we switch from OOP to IP polarizations, 2 situations can emerge: torsion and buckling. When the sample polarization is perpendicular to the cantilever main axis, a torsional motion of the cantilever occurs due to the IP deformation of the sample that drags the tip. This results in a horizontal photodiode signal (Fig. S4b). On the other hand, when the sample polarization is aligned with the cantilever long axis, the tip is dragged by the sample parallel to the axis, causing the buckling oscillation of the cantilever (Fig. S4c).

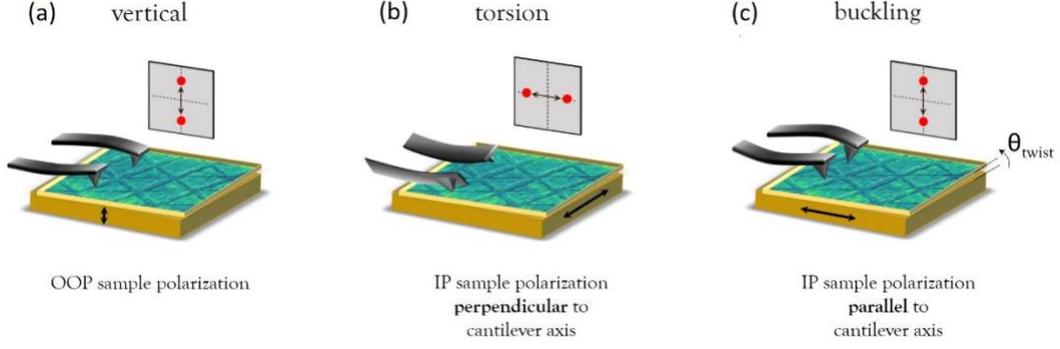

***Fig. S4.*** *(a) Schematic of OOP sample polarization causing vertical deflection of the cantilever, i.e. vertical movement of laser spot on the photodiode. (b) Representation of the torsional case, where an IP sample polarization is perpendicular to cantilever long axis. A lateral (horizontal) movement of laser spot is read on the photodiode. (c) Buckling case characterized by an IP sample polarization parallel to the cantilever axis. As in (a), the photodiode reads a vertical signal.*

As in the first case, a vertical motion of the laser spot onto the photodiode is observed. For this reason, the buckling effect is also known as an in-plane flexural crosstalk.

**Section 6: Saddle point OOP polarizations**

Here, we prove It is reasonable to consider that saddle point OOP polarizations responsible for the Fig. 2b phase contrast emerge from OOP PZ rather than FLX.

The general constitutive equation for the electric polarization $P_i$ reads:[20]

$$P_i = \chi_{ij} E_j + e_{ijk} u_{jk} + \mu_{klij} \frac{\partial u_{kl}}{\partial x_j} \tag{S4}$$

where $E_j$, $u_{jk}$ and $\frac{\partial u_{kl}}{\partial x_j}$ are the electric field, the strain tensor, and its spatial gradient, respectively. Einstein summation is adopted.[20] Spatial dependence of each term is omitted for simplicity.

Eq. S4 describes the appearance of a polarization field in a medium due to three phenomena. The first term describes the dielectric response according to the susceptibility tensor $\chi_{ij}$. The second provides the PZ response *via* the PZ tensor $e_{ijk}$. The third is related to the FLX polarization response to a strain gradient, through the FLX tensor $\mu_{klij}$.

A first distinction between PZ and FLX can be noticed by looking at the definition of the strain tensor $u_{jk}$. This is defined as the symmetric part of the tensor $\frac{\partial U_j}{\partial x_k}$, where $U_j$ is the local displacement of the point $x_j$ in the sample (with respect to the un-strained configuration):[20]

$$u_{jk} = \frac{1}{2}\left(\frac{\partial U_j}{\partial x_k} + \frac{\partial U_k}{\partial x_j}\right). \qquad (S5)$$

For the 2d (*x*, *z*) case, the PZ OOP polarization becomes proportional to the first derivative (along *x*) of the topography profile (*z* axis), while the FLX OOP polarization follows its second derivative. E.g., Fig. S5 plots a saddle point region, identified by the *x*-dependent OOP local deformation sketched in Fig. S5a. This is a reliable representation, as demonstrated by Fig. 6b. Accordingly, the first derivative (linked to the PZ polarization) and second derivative (linked to the FLX polarization) of the topography profile can be obtained. Fig. S5b displays the PZ OOP polarization: *two* main regions are visible where the polarization reaches maximum and minimum values. In Fig. S5c, instead, the FLX polarization is shown, characterized by *three* main local extrema.

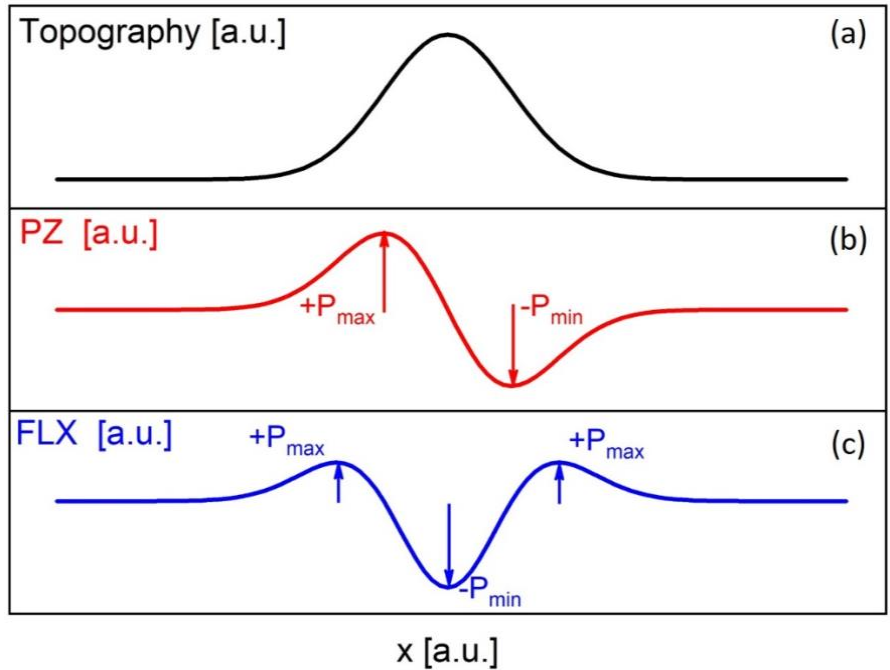

***Fig. S5.*** *(a) Topographical sketch of a saddle point region (see also Fig. 6b). (b) PZ OOP polarization profile proportional to the first derivative of the topography. (c) FLX OOP polarization profile obtained through topography second derivative. In (b, c), the positions of local maxima and minima of polarizations are highlighted by arrows.*

According to Fig. 2b, this supports the conclusion that the OOP polarization has a PZ nature. In Fig. 2b, only *two* extrema are visible along a profile crossing the saddle points (inset of Fig. 2b), one bright (~ -10°) and one dark (~ -30°). In contrast, if FLX was responsible for this OOP signal, we should expect *three* local extrema.

An additional consideration about the order of magnitudes of the involved polarization dipoles is needed, since one could question why OOP PZ is providing a PFM signal at the saddle points of the triangular moiré domains but not in the internal area. A possible argument is based on the proportionality of the different local polarizations to the dipole charges displacement. According to Ref. 19 and assuming that the internal area OOP polarization originates from a pure electronic contribution, the (vertical) displacement of those dipole charges - *d* - can be estimated to be ~0.1 pm. On the other hand, since saddle point OOP PZ dipoles are proportional to the slope of the topography, the order of magnitude for the distance *d* is the same of the topographical change across the saddle point (*i.e.*, *Δz* in Fig. 6b). This is at least one order of magnitude larger than 0.1 pm (Fig.6b), depending on $\theta_{TW}$. As a result, OOP polarizations at the saddle points should provide a stronger PFM signal compared to OOP polarizations localized in the internal area of the AB/BA moiré domains.

## Section 7: Twist angle extraction from PFM images

As mentioned in the main text, the actual $\theta_{TW}$ of a twisted LM may not be the same twist angle $\theta_{TW}$ at which two layers are superimposed. First, local imperfections of individual top and bottom LM flakes cause the formation of localized strain difference between the two, which alters the alignment of the atoms, causing a variable-size moiré pattern over the whole interface area between the twisted crystals. On top of this, at small $\theta_{TW} \leq 1°$, an additional displacement of the atoms is caused by atomic relaxation, which affects the saddle points regions.[3] Therefore, higher strain is expected at those regions. Simulations involving density functional theory and molecular dynamics are required to compute the new equilibrium position of the displaced atoms in the spatially varying stacking configuration imposed by the layer twisting.[3] However, this does not affect the overall periodicity of the moiré pattern. After rigid twisting of the layers, the regions characterized by a local AA stacking are not influenced by atomic relaxation.[21] Hence, it is possible to ignore atomic relaxation if the aim is to retrieve the real local $\theta_{TW}$.

The procedure to extract $\theta_{TW}$(and strain tensor) from an experimental image of a spatially varying moiré superlattice was described in Ref. 24. We report here the main results that we implement in our derivation of $\theta_{TW}$ to support Fig. 3 of the main text.

Let us consider two different 1L-hexagonal lattices vertically stacked with a relative in-plane rotation $\theta$ as in Fig. S6, with unit cell vectors of lengths $a_1$ and $a_2$, respectively. The orientations of the two lattices are referred to the $x$-axis. We assume that the vector connecting neighbouring A sites of the bottom lattice (red arrow) forms an angle $\phi_0$ (measured counter-clockwise). By defining the following unit vectors (unit vectors for a hexagonal lattice)

$$\boldsymbol{b_1} = \begin{pmatrix} \cos \phi_0 \\ \sin \phi_0 \end{pmatrix}, \qquad \boldsymbol{b_2} = \begin{pmatrix} \cos \left( \phi_0 + \frac{\pi}{3} \right) \\ \sin \left( \phi_0 + \frac{\pi}{3} \right) \end{pmatrix},$$

we can write the positions of all the A sites of the bottom layer (blue-shaded circles) as

$$\boldsymbol{r_A^b} = a_1 [\boldsymbol{b_1} \quad \boldsymbol{b_2}] \begin{pmatrix} m_1 \\ n_1 \end{pmatrix},$$

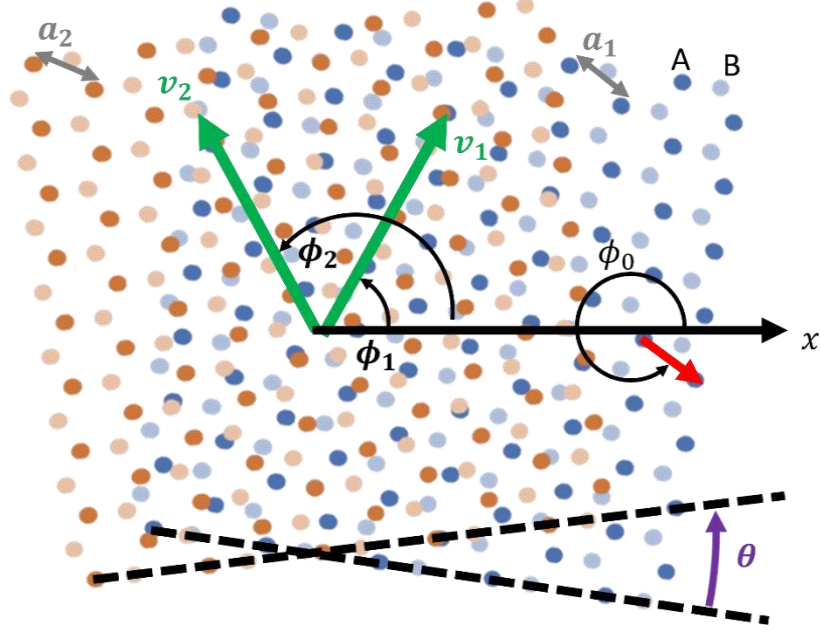

***Fig. S6.*** *Schematic representation of a moiré superlattice emerging from two hexagonal lattices (orange and blue shaded dots representing the A and B sites of the crystal) twisted by an in-plane angle θ. The two lattices have different lattice parameters $a_1, a_2$ (grey). Green arrows: moiré unit cell vectors $v_1, v_2$. Red arrow: primitive vector of bottom lattice (blue-shaded dots) connecting neighbouring A sites. All the angles measured counter-clockwise with respect to the x-axis.*

where $m_1, n_1$ are integers (representing the number of lattice sites away from the origin, which is set on one of the A sites). The same can be done for the position of the A sites in the top layer, by considering the inter-layer rotation (through the rotation matrix $R_\theta = \begin{pmatrix} \cos\theta & -\sin\theta \\ \sin\theta & \cos\theta \end{pmatrix}$) and the presence of a position-dependent displacement ($\hat{\Sigma} r$, where $\hat{\Sigma}$ is the strain matrix):

$$\boldsymbol{r}_A^t = R_\theta a_2 [\boldsymbol{b}_1 \quad \boldsymbol{b}_2] \begin{pmatrix} m_2 \\ n_2 \end{pmatrix} + \hat{\Sigma} \boldsymbol{r},$$

for integer values of $m_2, n_2$. In this case, all the strain (assumed to be uniform) is encoded in the top layer, while the bottom lattice is assumed to be non-deformed. This does not change the results, as the meaningful quantity is the strain difference between top and bottom layers, which alters the atoms alignment, in contrast to a common strain term.[22] Given the above definitions, it is possible to express the lattice sites of the moiré superlattice in terms of $r_A^b$ and $r_A^t$. Due to the hexagonal shape of the two layers, the emerging moiré pattern is also characterized by a hexagonal lattice with unit vectors:[22]

$v_1 = \lambda_1(\cos\phi_1, \sin\phi_1)$ and $v_2 = \lambda_2(\cos\phi_2, \sin\phi_2)$,

with orientations measured with respect to the $x$-axis. In absence of strain, $\lambda_1 = \lambda_2$ as there is no asymmetry in the two lattices, however, in general, the two values are different. According to Ref.24, an expression for the moiré lattice vectors is:

$$[v_1 \quad v_2] = a_2\left(R_{-\theta}(I - \hat{\Sigma}) - (1+\delta)I\right)^{-1}[b_1 \quad b_2]$$

where $I$ is the 2x2 identity matrix and $\delta = \frac{a_2}{a_1} - 1$. By rewriting this, we end up with an expression for the strain matrix in terms of $\theta, \phi_0$, and the experimentally observable parameters $\lambda_1, \lambda_2, \phi_1, \phi_2$:

$$\hat{\Sigma} = I - R_\theta(a_2[b_1 \quad b_2][v_1 \quad v_2]^{-1} + (1+\delta)I).$$

By imposing that $\hat{\Sigma}$ is $\theta$-independent (*i.e.*, $\hat{\Sigma}_{12} = \hat{\Sigma}_{21}$), as we are neglecting any atomic relaxation term,[22] we find the parametric relationship ($\phi_0$ is the parameter) between $\theta_{TW}$ and the observable parameters $\lambda_1, \lambda_2, \phi_1, \phi_2$:

$$e^{i\theta(\phi_0)} = \frac{1}{R(\phi_0)}(x(\phi_0) + iy(\phi_0)) \tag{S6}$$

with:

$$x(\phi_0) = x_0 + r_- \cos(\alpha_- - \phi_0)$$

$$y(\phi_0) = r_- \sin(\alpha_- - \phi_0)$$

$$R(\phi_0) = \sqrt{x(\phi_0)^2 + y(\phi_0)^2}$$

Where:

$$\Delta\phi = \phi_2 - \phi_1,$$

$$x_0 = \frac{2(1+\delta)}{a_2}\lambda_1\lambda_2 \sin\Delta\phi,$$

$$r_\pm = \sqrt{\lambda_1^2 + \lambda_2^2 - 2\lambda_1\lambda_2 \cos\left(\Delta\phi \mp \frac{\pi}{3}\right)},$$

$$e^{i\alpha_\pm} = \frac{i}{r_\pm}\left(\lambda_1 e^{i\left(\phi_1 \pm \frac{\pi}{3}\right)} - \lambda_2 e^{i\phi_2}\right).$$

As a result, $\hat{\Sigma}$ is a symmetric matrix that can be written as:

$$\hat{\Sigma} = \epsilon_c I + \epsilon_s \begin{pmatrix} \cos\gamma & \sin\gamma \\ \sin\gamma & -\cos\gamma \end{pmatrix}$$

where $\epsilon_c$ and $\epsilon_s$ represent the isotropic compression and shear strain terms, respectively, and $\gamma$ defines the strain direction:

$$\epsilon_s = \frac{a_2 r_+}{2\lambda_1\lambda_2 \sin\Delta\phi} \tag{S7}$$

$$\epsilon_c(\phi_0) = 1 - \frac{1+\delta}{x_0} R(\phi_0) \tag{S8}$$

$$e^{i\gamma(\phi_0)} = e^{i(\theta(\phi_0) + \alpha_+ + \phi_0 + \pi)}. \tag{S9}$$

Eqs. S6-S9 allow us to extract information about the (local) θ_TW and strain values from the knowledge of geometrical quantities of a moiré pattern, *i.e.*, size and shape of superlattice. However, as emerges from the parametric dependence of $\theta$, $\epsilon_c$ and $\gamma$, this assumes $\phi_0$ is known, otherwise multiple solutions can exist. In most cases the exact orientation of the bottom layer is not known. Nevertheless, a practical way to choose a proper value for $\phi_0$ is the one that minimizes $\epsilon_c^2$ (as suggested in Ref. 24). As most of LM crystals have an in-plane hexagonal lattice, the described framework is rather general, however, here we focus on the extraction of θ_TW le and strain from an experimental image of a t-hBN moiré superlattice: hence we set $a_1 = a_2 = a = 0.25$ nm and $\delta = 0$.

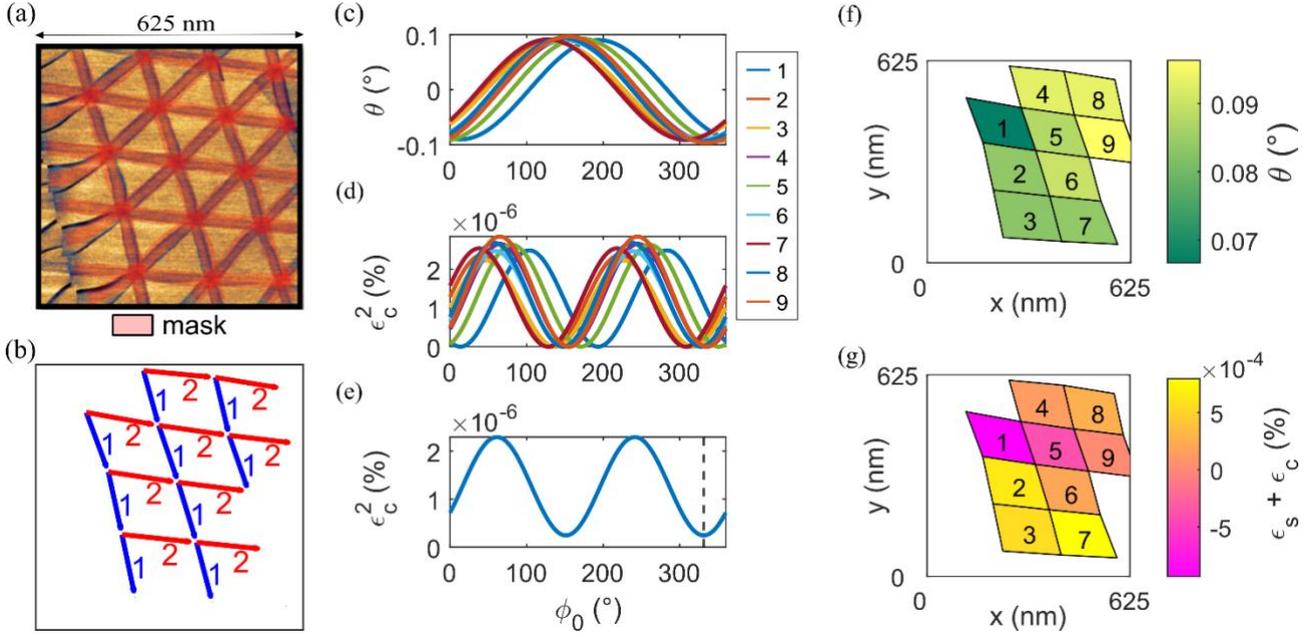

*Fig. S7.* Procedure for the extraction of local $\theta_{TW}$ and strain from an image of a moiré pattern. (a) PFM map of t-hBN moiré pattern (Fig. 3c of the main text) with a mask to filter for the perimeter of each triangular moiré domain. (b) Local lattice vectors (1 and 2) corresponding to each moiré domain fully enclosed in the image (9 domains are defined this way). Values of $\theta_{TW}$ (c) and square of the compressive strain (d) as a function of bottom lattice orientation $\phi_0$ for each moiré domain defined by the lattice vectors in (b). (d) Average $\epsilon_c^2$ on all 9 domains (labelled in (f), (g)). Dashed line indicates the minimum (in the $\pi$-$2\pi$ range) that fixes the value for $\phi_0$. (f, g) Map of $\theta_{TW}$ and total strain, respectively, corresponding to the image in (a), showing the position-dependent values.

In real-case scenarios, measured moiré patterns are irregular across a sufficiently large (tens of unit cells) detection area, *i.e.*, $v_1, v_2$ are functions of the position, hence also the strain matrix and twist angle. Therefore, all the above formulas are valid only locally (assuming an infinite pattern with the same local size and orientation).

To retrieve $\theta$,[22] our starting point is a typical PFM map, as in Fig. S7a. The triangular moiré domains change across the image, meaning that Eqs. S6-S9 hold for each triangle individually, which in turns has different values of $\theta_{TW}$ and strain. We first apply a mask (using the edge detection function in Gwyddion)[23] to isolate the perimeter of each triangular domain of the measured moiré superlattice.

From this mask, we can extract the intersection points representing the AA stacking regions (this is done with the help of a Radon transform in MATLAB, which allows for the identification of the direction of each line of the mask)[24]. The distance from neighbouring AA sites defines the moiré lattice vectors for each domain, as shown in Fig. S7b, where all the extracted couples ($\boldsymbol{v_1}, \boldsymbol{v_2}$) are shown (we consider only AA sites for which the neighbouring sites fall within the image boundaries, 9 in total). These provide the experimental (local) values for $\lambda_1, \lambda_2, \phi_1, \phi_2$. For each of the 9 moiré unit cells defined by the vectors in (b), we compute $\theta$ and $\epsilon_c^2$ as a function of the unknown parameter $\phi_0$, see Fig. S7c, d. These functions are slightly different because of the variation of the $\boldsymbol{v_1}, \boldsymbol{v_2}$ vectors in the image. We fix $\phi_0$ by minimizing the average $\epsilon_c^2$ for the entire image, shown by the dashed line in Fig. S7e (in doing this, $\epsilon_s$ is ignored since is $\phi_0$-independent). The resulting values for the extracted $\theta_{TW}$ and total strain as a function of position in the imaged area are shown in Fig. S7f, g, respectively. Each moiré cell (uniform since atomic relaxation is ignored) is characterized by the corresponding $\theta$ and $\epsilon_s + \epsilon_c$ at the target $\phi_0$.

The main result is that the values for $\theta$ are very close to $\theta_{TW}$ derived for an unstrained lattice: the average value for Fig. S7f is 0.086°, which only differs by 7% from 0.093° for the unstrained case. We repeated the same extraction procedure also for the other images shown in Fig. 3, finding differences no larger than 7%, thus validating the values reported in the main text.

Regarding the total strain, instead, we have very small values compared to what is found, e.g. in Ref. 24 for a moiré pattern that presents a morphological deformation caused by the fabrication process (a bubble). This is due to the flatness of the scanned areas in Fig. 3, for which the strain is ≈0.0001%. This approach does not consider atomic relaxation (the displacement is assumed continuously varying through the term $\hat{\Sigma}\boldsymbol{r}$), thereby the real value for the total amount of strain may be underestimated.[25]

**Section 8: Shape evolution of two anti-parallel stacking moiré superlattices**

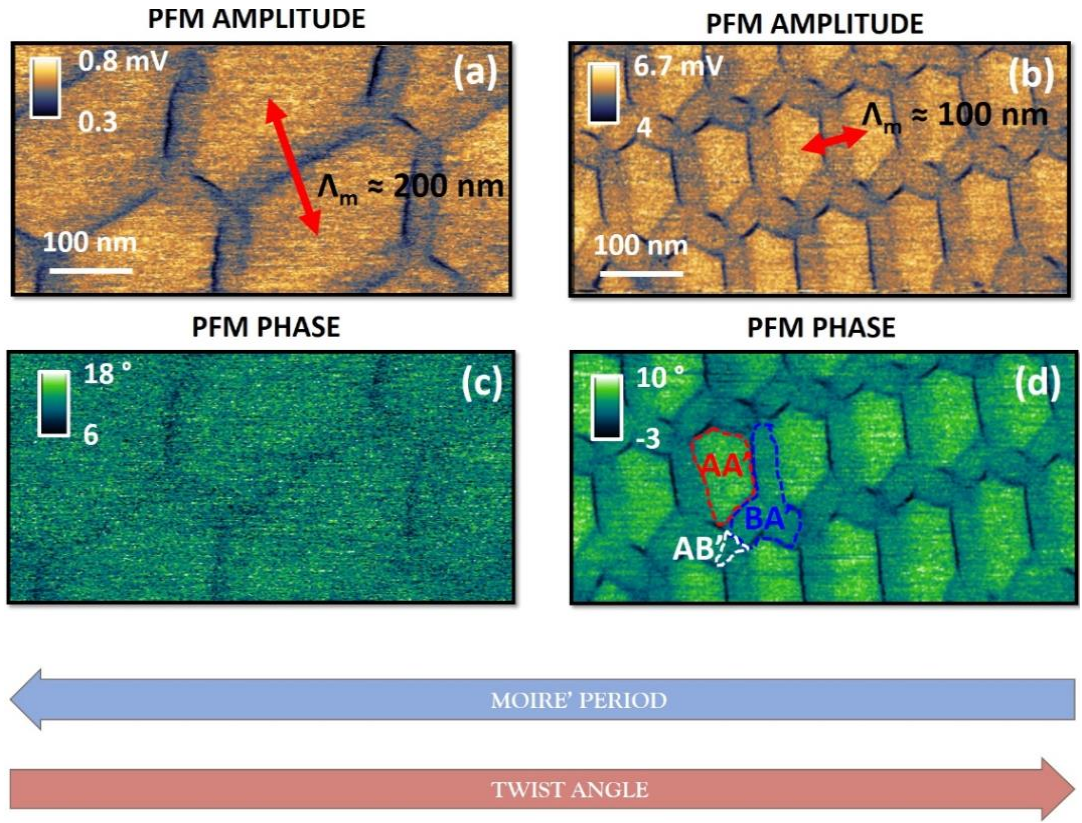

*Fig. S8.* *(a, c) PFM amplitude and phase images of anti-parallel moiré superlattice with Λ$_m$~200 nm (red arrow). (b, d) PFM amplitude and phase maps of smaller moiré superlattice with Λ$_m$~100 nm (red arrow). In (d) the positions of AA' (red), BA' (blue) and AB' (white) stacking are indicated.*

Fig.S8 shows two moiré superlattices of an anti-parallel t-hBN sample (top flake≈4.5nm, bottom layer thickness~40 nm, $\theta_{TW}$~0.2°). These two regions are characterized by a different local $\theta_{TW}$, therefore, Λ$_m$, due to fabrication imperfections. Both PFM amplitude and phase maps display moiré patterns typical of anti-parallel stacking, confirming that AA' hexagonal domains occupy most of the superlattice, as they are energetically favoured (Fig. S1). Figs. S8a, c show PFM signals from a region where Λ$_m$~200 nm. Figs. S8b, d plot the PFM signals of a different region of the same t-hBN, where the moiré superlattice is characterized by a smaller Λ$_m$~100 nm. As observed in Fig. 3, if Λ$_m$ gets smaller (from Fig. S8a to b), the relative area covered by the less energetically favoured domains (here, AB'/BA') should increase (opposite trend for AA' hexagonal regions). This is seen in Fig. S8.

**Section 9: PFM amplitude images for Fig. 4 of main text**

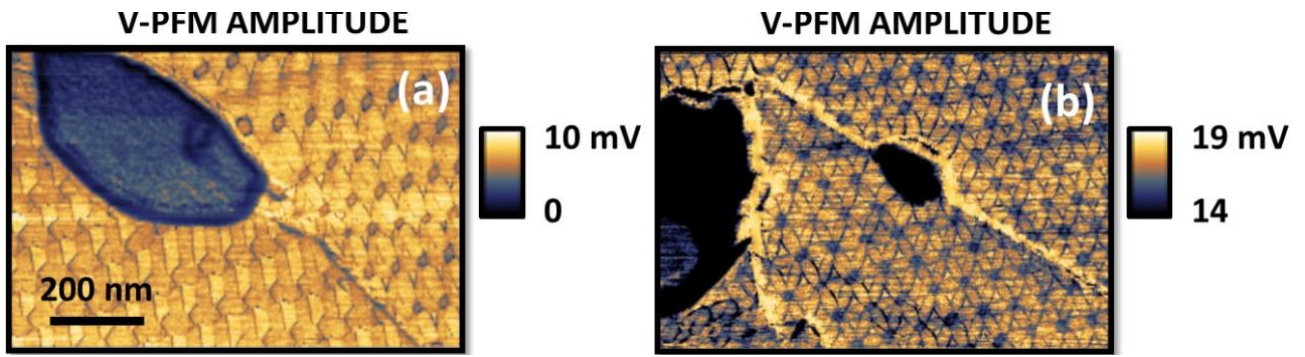

*Fig. S9.* *(a) PFM amplitude map related to PFM phase map of Fig. 4b. (b) PFM amplitude image related to PFM phase map of Fig. 4g.*

**Section 10: Schematic of sample showing a double-moiré in Fig. 5a.**

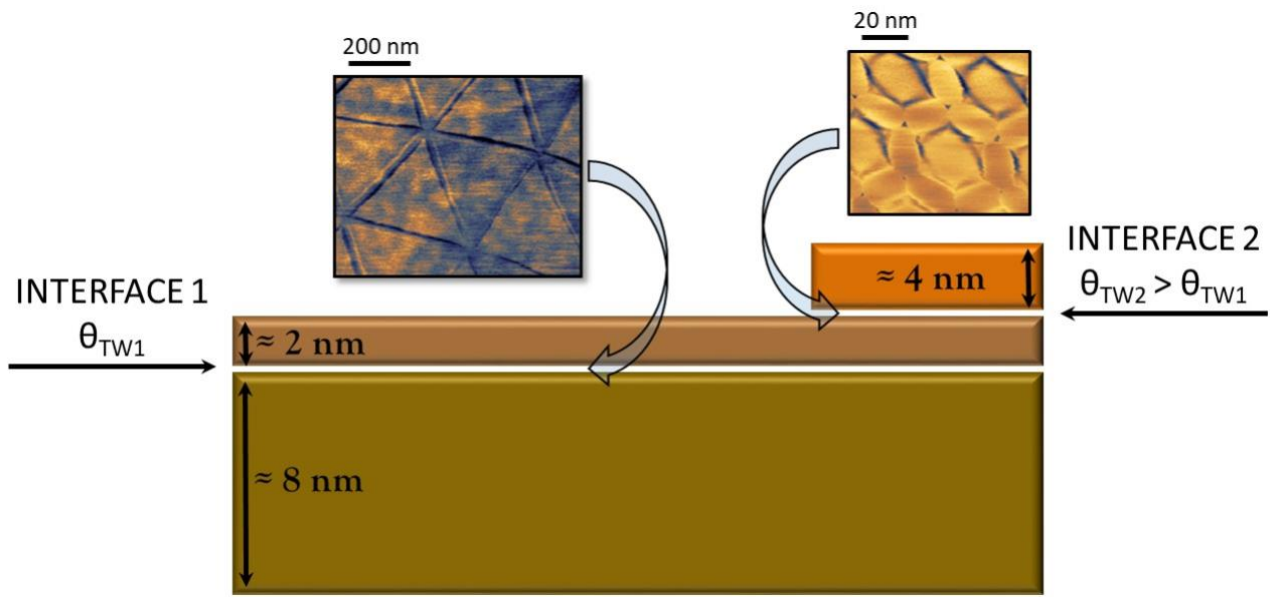

*Fig. S10.* *Schematic t-hBN sample providing a double-moiré in Fig. 5. 3 hBN flakes are present with two interfaces (1 and 2). Correspondingly, two moiré superlattices are present. At "interface 2" they overlap, providing a double-moiré pattern.*

**Section 11: PFM topographical image of a t-hBN moiré pattern**

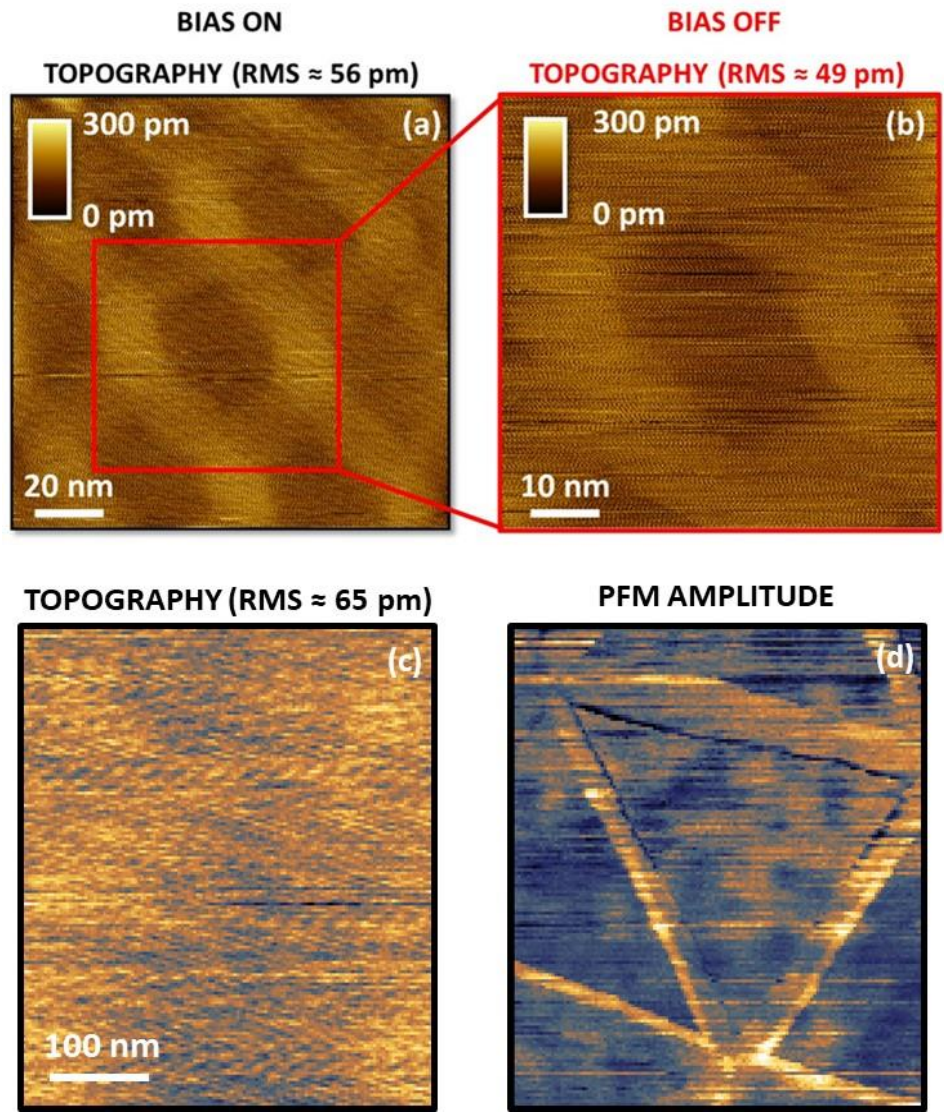

*Figure S11.* PFM contact mode topography of a t-hBN moiré pattern. (a) Topography image obtained scanning the sample in PFM with $V_{dc} = 0$ V, $V_{ac} = 2V$. The RMS of the image is on top of the panel. (b) Topography zoomed image achieved by scanning the sample in PFM without any applied DC or AC voltage. The RMS of the image is shown on top of the panel. (c) Topography of a different t-hBN region showing a similar RMS to the one of panel (a) and (b). (d) PFM amplitude image corresponding to panel (c), confirming the presence of a moiré triangular domain with a bigger length, $\Lambda_m \sim 250$ nm, than in panel (a), $\Lambda_m \sim 55$ nm.

Here, we prove the topographical image shown in Fig. 6a (reported in Fig. S11a) not to emerge from any bias related artefact. For this purpose, we scann the same area first in standard PFM (Fig. S10a), applying a bias between tip and sample ($V_{dc} = 0$ V, $V_{ac} = 2$V), and then in standard PFM but with $V_{ac}=0$ (Fig. S11b). As shown in Fig. S11, in both cases the topography follows the moiré pattern structure, highlighting an intrinsic modulation of the hBN surface, not influenced by the presence of a PFM bias. Additionally, we do PFM measurements of a different t-hBN region showing a triangular moiré pattern with $\Lambda_m \sim 250$ nm (much bigger than $\Lambda_m \sim 55$ nm, as for Fig. S11a). As can be seen in Fig. S11 c, the topography does not show any moiré feature. Since the topographical RMS of Fig. S11a and S11c are very similar, we can exclude that in Fig. S11c the topographical moiré pattern is not visible due to a higher image noise level.